\documentclass[superscriptaddress,preprintnumbers]{revtex4}

\usepackage{graphicx}
\newcommand*{\ie}{{\it i.e.},\ }
\newcommand*{\etal}{{\it et al.\ }}
\newcommand*{\gev}{{\rm\,GeV}}
\newcommand*{\gevc}{{\rm\,GeV/c}}
\newcommand*{\tev}{{\rm\,TeV}}
\newcommand*{\pt}{$p_T$}

\newcommand*{\UE}{``underlying event"}
\newcommand*{\BBR}{``beam-beam remnants"}
\newcommand*{\MB}{Min-Bias}
\newcommand*{\pthard}{$p_T({\rm hard})$}
\newcommand*{\ptcut}{$p_T\!>\!0.5\,{\rm GeV/c}$}
\newcommand*{\etacut}{$|\eta|\!<\!1$}
\newcommand*{\hardcut}{$p_T\!({\rm hard})>3{\rm\,GeV/c}$}
\newcommand*{\hardzero}{$p_T\!({\rm hard})>0{\rm\,GeV/c}$}
\newcommand*{\ptchj}{$P_T\!({\rm chgjet}\#1)$}
\newcommand*{\aveN}{$\langle\! N_{\rm chg}\!\rangle$}
\newcommand*{\avePT}{$\langle P_T{\rm sum}\!\rangle$}
\newcommand*{\etaphi}{$\eta$-$\phi$}
\begin{document}
% You should use BibTeX and revtex.bst for references
\bibliographystyle{revtex}

% number  and your conference paper identification number on the
% title page in preprint mode. Multiple \preprint commands are allowed.
\preprint{CDF/MIN-BIAS/PUBLIC/5746}
\preprint{UFIFT-HEP-01-19}

\title{The Underlying Event in Hard Scattering Processes}
% Optional argument for running titles on pages
%\title[]{}



\author{Rick Field}
\email[Talk presented at Snowmass 2001 (P5 working group);\ ]{rfield@phys.ufl.edu}
\homepage[]{http://www.phys.ufl.edu/~rfield/}
%\thanks{}
\affiliation{
Department of Physics, University of Florida,
Gainesville, Florida, 32611, USA}

\collaboration{for the CDF Collaboration}
\noaffiliation

\date{\today}

\begin{abstract}
We study the behavior of the \UE\ in hard scattering proton-antiproton collisions at $1.8\tev$ and 
compare with the QCD Monte-Carlo models. The \UE\ is everything except the two outgoing hard 
scattered ``jets" and receives contributions from the \BBR\ plus initial and final-state radiation. The 
data indicate that neither ISAJET or HERWIG produce enough charged particles (with \ptcut) from the 
``beam-beam remnant" component and that ISAJET produces too many charged particles from initial-state radiation.  
PYTHIA which uses multiple parton scattering to enhance the \UE\ does the best job describing the 
data.
\end{abstract}
% \pacs{}

%\maketitle must follow title, authors, abstract and \pacs
\maketitle

%\section{}
%\label{}
%\subsection{}
%\subsubsection{}

\section{Introduction}

\begin{figure}[htbp]
\includegraphics[scale=0.8]{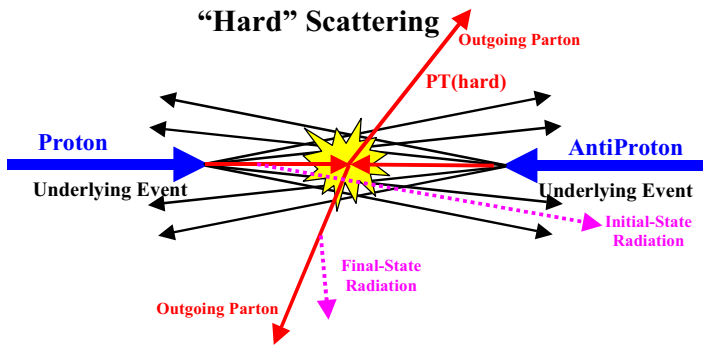}
\caption{Illustration of the way QCD Monte-Carlo models simulate a proton-antiproton collision in 
which a ``hard" $2$-to-$2$ parton scattering with transverse momentum, \pthard, has occurred.  The 
resulting event contains particles that originate from the two outgoing partons 
({\it plus initial and final-state radiation}) and particles that come from the breakup of the 
proton and antiproton ({\it\BBR}).  The \UE\ is everything except the two outgoing hard scattered ``jets" 
and consists of the \BBR\ plus initial and final-state radiation. The ``hard scattering" component 
consists of the outgoing two ``jets" plus initial and final-state radiation.
}
\label{snow_fig1}
\end{figure}

FIG.~\ref{snow_fig1} illustrates the way QCD Monte-Carlo models simulate a proton-antiproton collision in which 
a "hard" $2$-to-$2$ parton scattering with transverse momentum, \pthard, has occurred.  The resulting event contains 
particles that originate from the two outgoing partons ({\it plus initial and final-state radiation}) and particles 
that come from the breakup of the proton and antiproton (\ie \BBR).  The \UE\ is everything except the 
two outgoing hard scattered ``jets" and receives contributions from the \BBR\ plus initial and final-state radiation. 
The ``hard scattering" component consists of the outgoing two ``jets" plus initial and final-state radiation.

\begin{figure}[htbp]
\includegraphics[scale=0.8]{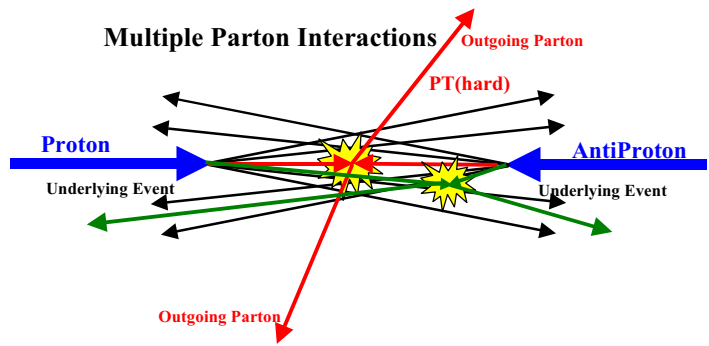}
\caption{Illustration of the way PYTHIA models the \UE\ in proton-antiproton collision by including multiple 
parton interactions. In adddition to the hard $2$-to-$2$ parton-parton scattering with transverse momentum, \pthard, 
there is a second ``semi-hard" $2$-to-$2$ parton-parton scattering that contributes particles to the \UE.
}
\label{snow_fig2}
\end{figure}

The \BBR\ are what is left over after a parton is knocked out of each of the initial two beam 
hadrons.  It is the reason hadron-hadron collisions are more ``messy" than electron-positron annihilations and no one 
really knows how it should be modeled.  For the QCD Monte-Carlo models the \BBR\ are an 
important component of the \UE.  Also, it is possible that multiple parton scattering contributes to the 
\UE.  FIG.~\ref{snow_fig2} shows the way PYTHIA \cite{pythia} models the \UE\ in proton-antiproton collision 
by including multiple parton interactions. In addition to the hard $2$-to-$2$ parton-parton scattering and the \BBR, 
sometimes there is a second ``semi-hard" $2$-to-$2$ parton-parton scattering that contributes particles to the \UE.  

Of course, from a certain point of view there is no such thing as an \UE\ in a proton-antiproton 
collision.  There is only an ``event" and one cannot say where a given particle in the event originated.  On the other 
hand, hard scattering collider ``jet" events have a distinct topology.  On the average, the outgoing hadrons 
``remember" the underlying the $2$-to-$2$ hard scattering subprocess.  An average hard scattering event consists of a 
collection (or burst) of hadrons traveling roughly in the direction of the initial beam particles and two collections of 
hadrons (\ie ``jets") with large transverse momentum.  The two large transverse momentum ``jets" are roughly back 
to back in azimuthal angle.  One can use the topological structure of hadron-hadron collisions to study the 
\UE\ \cite{field_prd,field_dpf,tano}. The ultimate goal is to understand the physics of the \UE, but since it is very 
complicated and involves both non-perturbative as well as perturbative QCD it seems unlikely that this will happen 
soon.  In the mean time, we would like to tune the QCD Monte-Carlo models to do a better job fitting the \UE.  
The \UE\ is an unavoidable background to most collider observables.  To find ``new" physics 
at a collider it is crucial to have Monte-Carlo models that simulate accurately ``ordinary" hard-scattering collider 
events.  In this talk I will compare collider observables that are sensitive to the \UE\ with the QCD 
Monte-Carlo model predictions of PYTHIA 6.115 \cite{pythia}, HERWIG 5.9 \cite{herwig}, and ISAJET 7.32 \cite{isajet}
and discuss the tuning of PYTHIA.

\begin{figure}[htbp]
\includegraphics[scale=0.8]{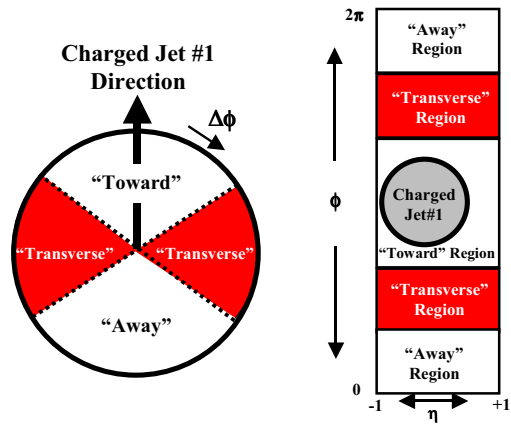}
\caption{Illustration of correlations in azimuthal angle $\Delta\phi$ relative to the direction of the 
leading charged jet in the event, chgjet\#1.  The angle $\Delta\phi=\phi-\phi_{\rm chgjet\#1}$  
is the relative azimuthal angle between charged particles and the direction of chgjet\#1.  The``toward" 
region is defined by $|\Delta\phi|<60^\circ$ and \etacut, while the ``away" 
region is $|\Delta\phi|>120^\circ$ and \etacut.   The ``transverse" region is defined by 
$60^\circ<|\Delta\phi|<120^\circ$  and \etacut.  Each region has an 
area in \etaphi\ space of $4\pi/3$.
}
\label{snow_fig3}
\end{figure}

\begin{figure}[htbp]
\includegraphics[scale=0.8]{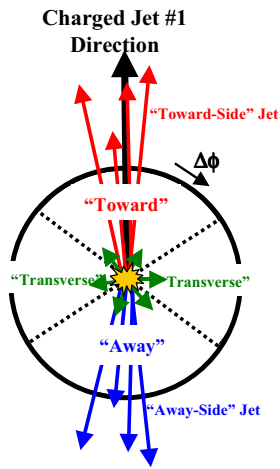}
\caption{Illustration of the topology of an average proton-antiproton collision in which a ``hard" 
$2$-to-$2$ parton scattering has occurred.  The ``toward" region as defined in FIG.~\ref{snow_fig3}
contains the leading charged particle ``jet", while the ``away" region, on the average, contains 
the ``away-side" jet.  The ``transverse" region is perpendicular to the plane of the hard $2$-to-$2$ 
scattering and is very sensitive to the \UE.
}
\label{snow_fig4}
\end{figure}

\section{The ``Transverse" Region}

In a proton-antiproton collision large transverse momentum outgoing partons manifest themselves, in the laboratory, 
as a clusters of particles ({\it both charged and neutral}) traveling in roughly the same direction.  These clusters are 
referred to as ``jets".  In this analysis we examine only the charged particle component of ``jets".  Our philosophy in 
comparing the QCD Monte-Carlo models with data is to select a region where the data is very ``clean" so that ``what 
you see is what you get" ({\it almost}).  Hence, we consider only charged particles measured by the CDF central tracking 
chamber (CTC) in the region \ptcut\ and \etacut, where the track finding efficiency is high and uniform 
(estimated to be $92\%$ efficient) and we restrict ourselves to charged particle jets with transverse momentum less than 
$50\gevc$.  The data presented here are uncorrected.  Instead the theoretical Monte-Carlo models are corrected for 
the track finding efficiency by removing, on the average, $8\%$ of the charged particles.  The theory curves have an 
error ({\em statistical plus systematic}) of about $5\%$.   Thus, to within $10\%$ ``what you see is what you get".

\begin{figure}[htbp]
\includegraphics[scale=0.6]{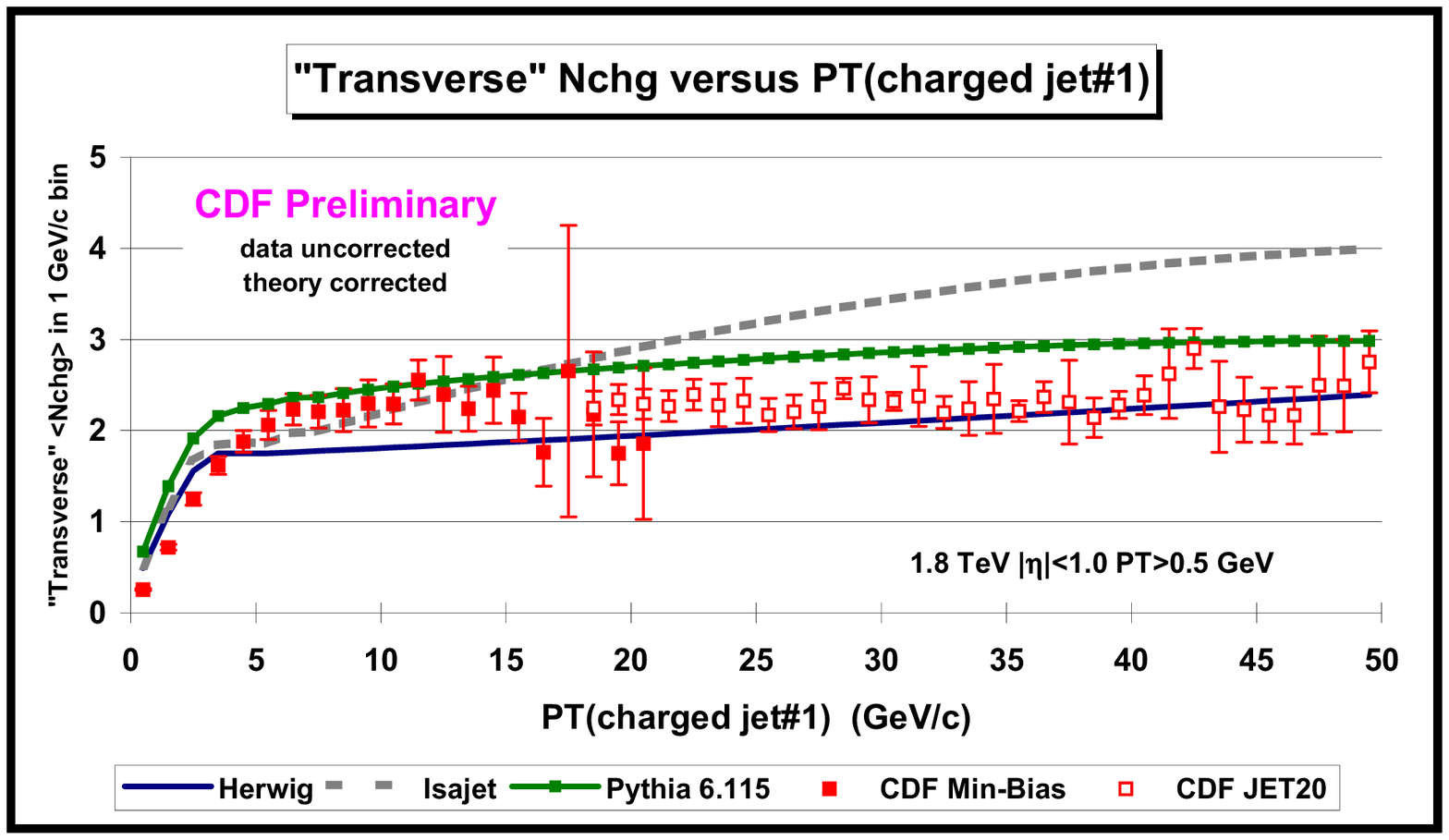}
\caption{Data on the average number of charged particles (\ptcut, \etacut) in the ``transverse" region defined in 
FIG.~\ref{snow_fig3} as a function of transverse momentum of the leading charged jet compared with the QCD 
Monte-Carlo predictions of HERWIG 5.9, ISAJET 7.32, and PYTHIA 6.115 with their default parameters and 
with \hardcut. Each point corresponds to the \aveN\  in a $1\gevc$ bin. The solid (open) points are the 
\MB\ (JET20) data. The theory curves are corrected for the track finding efficiency and have an 
error ({\it statistical plus systematic}) of around $5\%$.
}
\label{snow_fig5}
\end{figure}

\begin{figure}[htbp]
\includegraphics[scale=0.6]{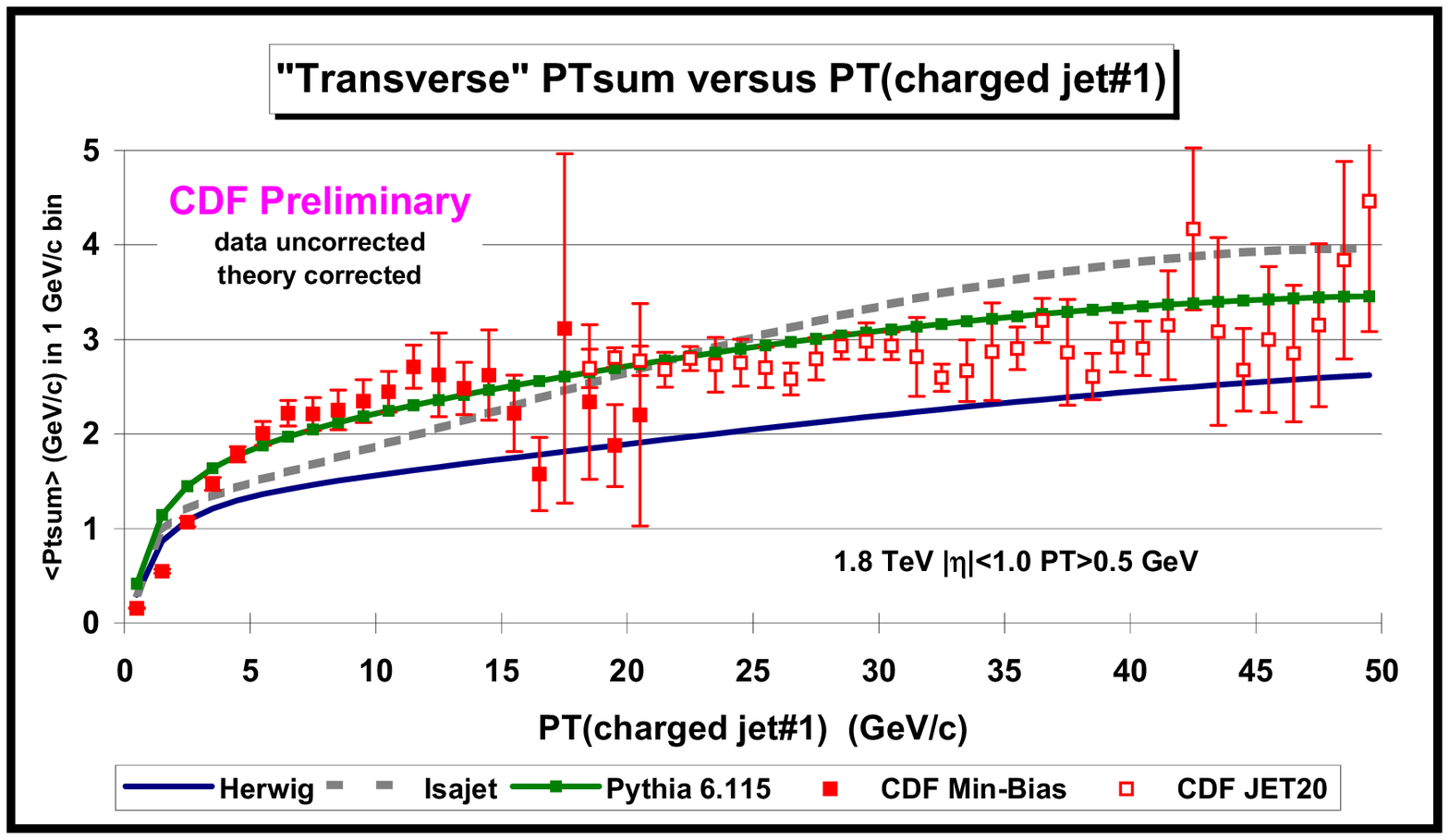}
\caption{Data on the average {\it scalar} \pt\ sum of charged particles (\ptcut, \etacut) in the ``transverse" 
region defined in FIG.~\ref{snow_fig3} as a function of the transverse momentum of the leading charged jet 
compared with the QCD Monte-Carlo predictions of HERWIG 5.9, ISAJET 7.32, and PYTHIA 6.115 with their default 
parameters and with \hardcut. Each point corresponds to the \avePT\ in a $1\gevc$ bin. The solid (open) points 
are the \MB\ (JET20) data.  The theory curves are corrected for the track finding efficiency and have an 
error ({\it statistical plus systematic}) of around $5\%$.
}
\label{snow_fig6}
\end{figure}

Charged particle ``jets" are defined as clusters of charged particles (\ptcut, \etacut) in ``circular regions" of 
\etaphi\ space with radius $R = 0.7$.   Every charged particle in the event is assigned to a ``jet", 
with the possibility that some jets might consist of just one charged particle.  The transverse momentum of a 
charged jet, $P_{T}\!({\rm chgjet})$, is the {\it scalar} \pt\ sum of the particles in the jet.  We use the 
direction of the leading charged particle jet to define correlations in azimuthal angle, $\Delta\phi$.  
The angle $\Delta\phi=\phi-\phi_{\rm chgjet\#1}$  is the relative azimuthal angle between a charged 
particle and the direction of chgjet\#1.  The``toward" region is defined by $|\Delta\phi|<60^\circ$ and \etacut, 
while the ``away" region is $|\Delta\phi|>120^\circ$ and \etacut.   The ``transverse" region is defined by 
$60^\circ<|\Delta\phi|<120^\circ$  and \etacut.  The three regions ``toward", ``transverse", and ``away" are 
shown in FIG.~\ref{snow_fig3}.  Each region has an area in \etaphi\ space of $4\pi/3$.  As illustrated in 
FIG.~\ref{snow_fig4}, the ``toward" region contains the leading charged particle jet, while the ``away" region, 
on the average, contains the ``away-side" jet.  The ``transverse" region is perpendicular to the plane of the 
hard $2$-to-$2$ scattering and is therefore very sensitive to the \UE. 

\begin{figure}[htbp]
\includegraphics[scale=0.6]{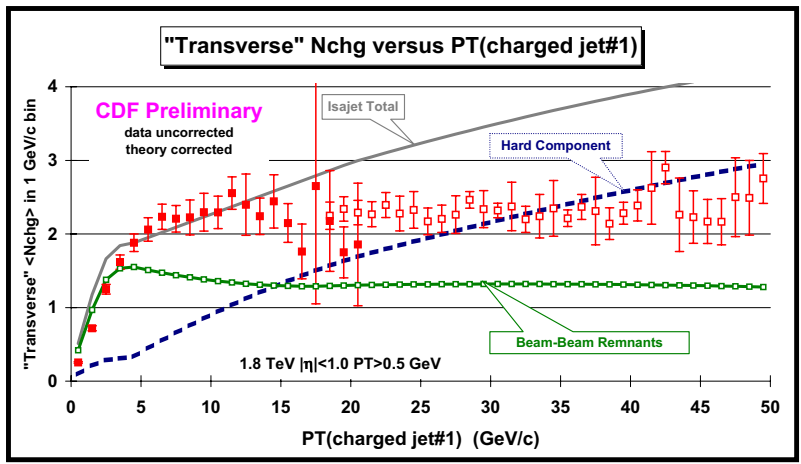}
\caption{Data on the average number of charged particles (\ptcut, \etacut) in the ``transverse" region defined 
in FIG.~\ref{snow_fig3} as a function of the transverse momentum of the leading charged jet compared with the 
QCD Monte-Carlo predictions of ISAJET 7.32 (default parameters and \hardcut). The predictions of ISAJET are divided 
into two categories: charged particles that arise from the break-up of the beam and target ({\it beam-beam remnants}), 
and charged particles that result from the outgoing jets plus initial and final-state radiation 
({\it hard scattering component}). The theory curves are corrected for the track finding efficiency and have an 
error ({\it statistical plus systematic}) of around $5\%$.
}
\label{snow_fig7}
\end{figure}

\begin{figure}[htbp]
\includegraphics[scale=0.6]{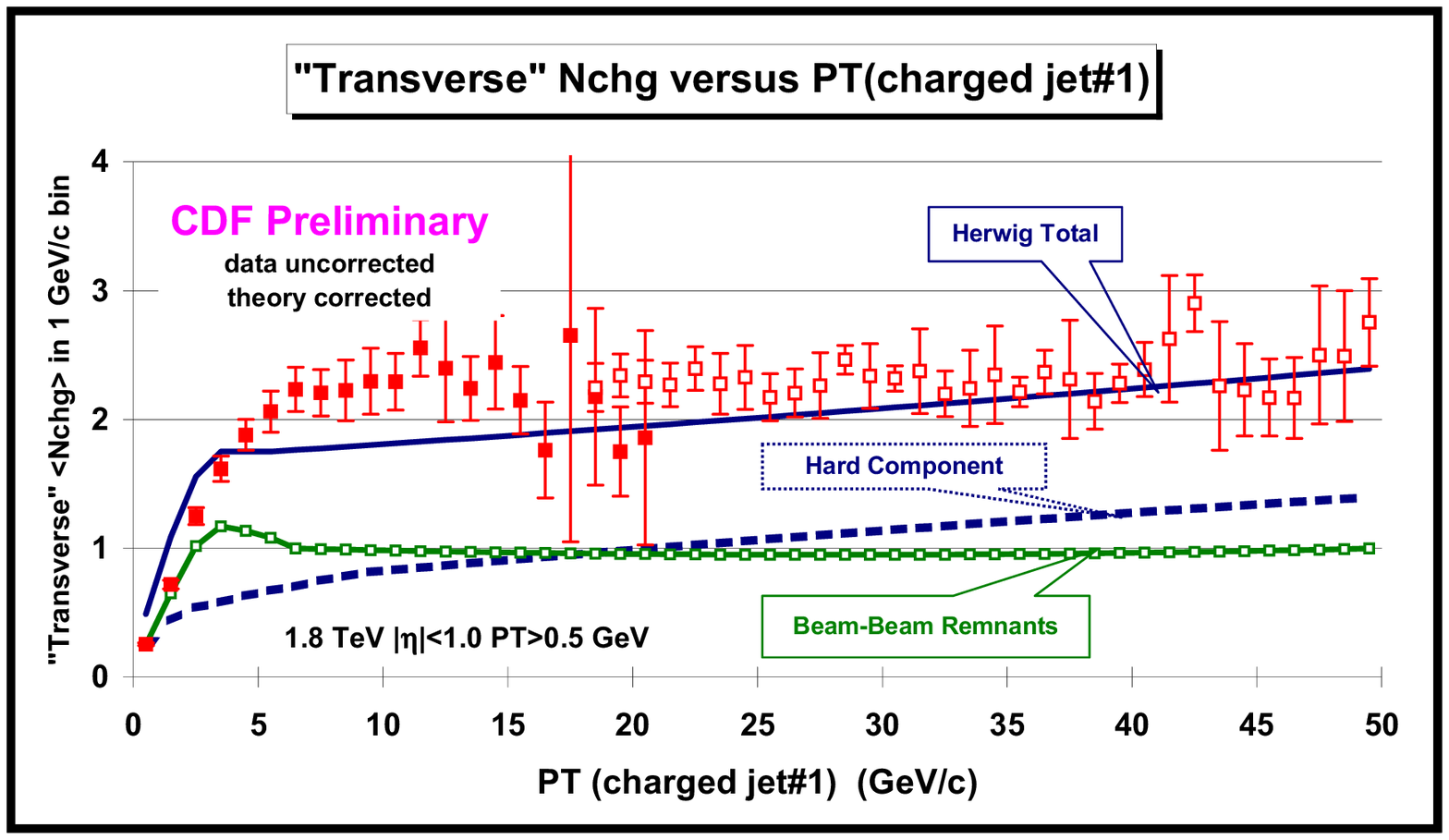}
\caption{Data on the average number of charged particles (\ptcut, \etacut) in the ``transverse" region defined 
in FIG.~\ref{snow_fig3} as a function of the transverse momentum of the leading charged jet compared with the 
QCD Monte-Carlo predictions of HERWIG 5.9 (default parameters and \hardcut). The predictions of HERWIG are divided 
into two categories: charged particles that arise from the break-up of the beam and target ({\it beam-beam remnants}), 
and charged particles that result from the outgoing jets plus initial and final-state radiation 
({\it hard scattering component}). The theory curves are corrected for the track finding efficiency and have an 
error ({\it statistical plus systematic}) of around $5\%$.
}
\label{snow_fig8}
\end{figure}

\begin{figure}[htbp]
\includegraphics[scale=0.6]{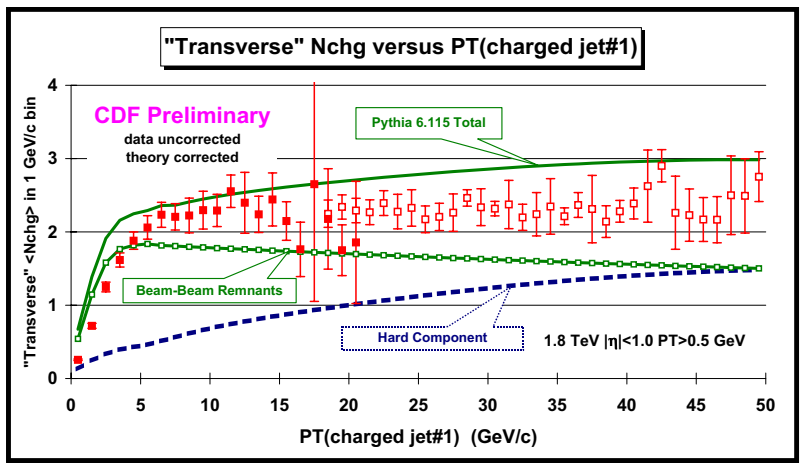}
\caption{Data on the average number of charged particles (\ptcut, \etacut) in the ``transverse" region defined 
in FIG.~\ref{snow_fig3} as a function of the transverse momentum of the leading charged jet compared with the 
QCD Monte-Carlo predictions of PYTHIA 6.115 (default parameters and \hardcut). The predictions of PYTHIA are divided 
into two categories: charged particles that arise from the break-up of the beam and target ({\it beam-beam remnants}), 
and charged particles that result from the outgoing jets plus initial and final-state radiation 
({\it hard scattering component}). For PYTHIA the \BBR\ include contributions from multiple parton 
scattering. The theory curves are corrected for the track finding efficiency and have an error 
({\it statistical plus systematic}) of around $5\%$.
}
\label{snow_fig9}
\end{figure}

\begin{figure}[htbp]
\includegraphics[scale=0.4]{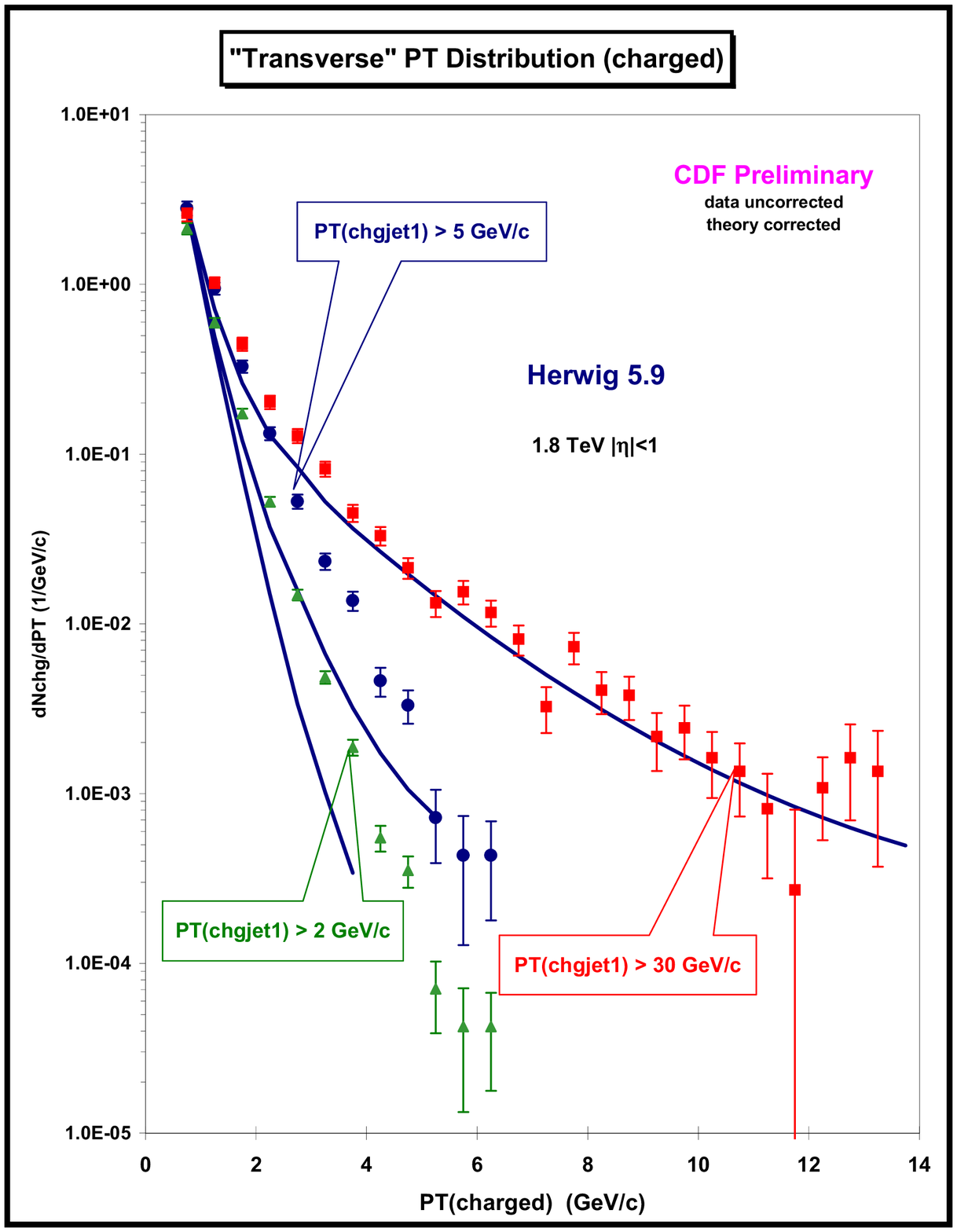}
\caption{Data on the transverse momentum distribution of charged particles (\ptcut, \etacut) in the ``transverse" region 
for \ptchj $>2\gevc$, $5\gevc$, and $30\gevc$, where chgjet\#1 is the leading charged particle jet. Each point 
corresponds to $dN_{chg}/dp_T$ and the integral of the distribution gives the average number of charged particles 
in the transverse region, $\langle N_{chg}({\rm transverse})\rangle$. The data are compared with the QCD Monte-Carlo 
model predictions of HERWIG 5.9 (default parameters and \hardcut).  The theory curves are corrected for the track 
finding efficiency and have an error ({\it statistical plus systematic}) of around $5\%$.
}
\label{snow_fig17}
\end{figure}

FIG.~\ref{snow_fig5} and FIG.~\ref{snow_fig6} compare the ``transverse" \aveN\ and the ``transverse" \avePT, 
respectively, with the QCD 
Monte-Carlo predictions of HERWIG, ISAJET, and PYTHIA 6.115 with their default parameters and \hardcut. The solid 
points are \MB\ data and the open points are the JET20 data. The JET20 data connect 
smoothly to the \MB\ data and allow us to study observables over the range $0.5 < P_T\!({\rm chgjet}\#1) < 50\gevc$.  
The average number of charged particles in the ``transverse" region doubles in going from \ptchj $=1.5\gevc$ to $2.5\gevc$ 
and then forms an approximately constant ``plateau" for \ptchj $>5\gevc$. If we 
suppose that the \UE\ is uniform in azimuthal angle $\phi$ and pseudo-rapidity $\eta$, the observed $2.3$ charged 
particles at \ptchj $=20\gevc$ translates to $3.8$ charged particles per unit pseudo-rapidity with \ptcut\
(multiply by $3$ to get $360^\circ$, divide by $2$ for the two units of pseudo-rapidity, 
multiply by $1.09$ to correct for the track 
finding efficiency).  We know that if we include all $p_T > 50$ MeV/c there are, on the average, about four charged 
particles per unit rapidity in a ``soft" proton-antiproton collision at 1.8 TeV \cite{CDF3}.  The data in 
FIG.~\ref{snow_fig5} imply that in the 
\UE\ of a hard scattering there are, on the average, about $3.8$ charged particles per unit rapidity with \ptcut!  
Assuming a charged particle \pt\ distribution of $e^{-2p_T}$ (see FIG.~\ref{snow_fig17}) implies that there are 
roughly $10$ 
charged particles per unit pseudo-rapidity with $p_T >  0$ in the \UE\ (factor of e).  Since we examine 
only those charge particles with \ptcut, we cannot accurately extrapolate to low \pt, however, it is clear that 
the \UE\ has a charge particle density that is at least a factor of two larger than the four charged 
particles per unit rapidity seen in ``soft" proton-antiproton collisions at this energy.  

The \MB\ data were collected with a very ``loose" trigger.  The CDF \MB\ trigger requirement removes 
elastic scattering and most of the single and double diffraction events, but keeps essentially all
the ``hard-scattering" events.  
In comparing with the QCD Monte-Carlo models we do require that the models satisfy the CDF \MB\ trigger, 
however, for \ptchj $> 5\gevc$ essentially all the generated events satisfy the \MB\ trigger (\ie the 
\MB\ trigger is unbiased for large \pt\ ``jets").  If we had enough \MB\ events we would not need the JET20 data, 
but because of the fast fall-off of the cross section we run out of statistics at around \ptchj $=20\gevc$ (that 
is why the \MB\ data errors get large at around $20\gevc$).  The JET20 data were collected by requiring at least 
$20\gev$ of energy ({\it charged plus neutral}) in a cluster of calorimeter cells.  We do not use the calorimeter 
information, but instead look only at the charged particles measured in the CTC in the same way we do for the \MB\
data.  The JET20 data is, of course, biased for low \pt\ jets and we do not show the JET20 data below 
\ptchj\ around $20\gevc$.  At large \ptchj\ values the JET20 data becomes unbiased and, in fact, we 
know this occurs at around $20\gevc$ because it is here that it agrees with the ({\it unbiased}) \MB\ data.  

\begin{figure}[htbp]
\includegraphics[scale=0.6]{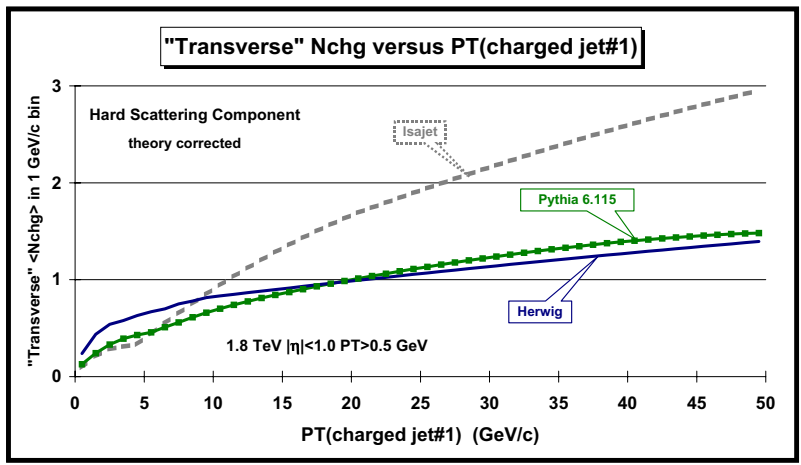}
\caption{The ``hard scattering" component ({\it outgoing jets plus initial and final-state radiation}) of the number 
of charged particles (\ptcut, \etacut) in the ``transverse" region defined in FIG.~\ref{snow_fig3} as a function of 
the transverse momentum of the leading charged jet from the QCD Monte-Carlo predictions of 
HERWIG 5.9, ISAJET 7.32, and PYTHIA 6.115 with their default parameters and with \hardcut. The curves are corrected 
for the track finding efficiency and have an error ({\it statistical plus systematic}) of around $5\%$.
}
\label{snow_fig10}
\end{figure}

\begin{figure}[htbp]
\includegraphics[scale=0.6]{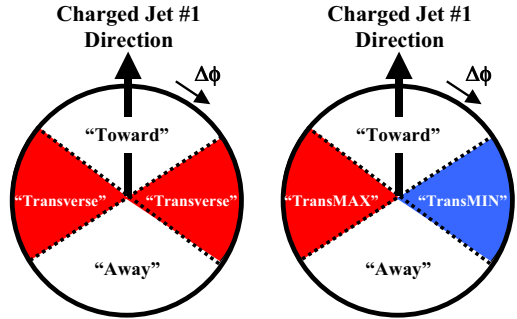}
\caption{Illustration of correlations in azimuthal angle $\Delta\phi$ relative to the direction of the leading charged 
jet in the event, chgjet\#1.  The angle $\Delta\phi=\phi-\phi_{\rm chgjet\#1}$ is the relative azimuthal angle between 
charged particles and the direction of chgjet\#1.  On an event by event basis, we define ``transMAX (``transMIN") to 
be the maximum (minimum) of the two ``transverse" pieces, $60^\circ<\Delta\phi<120^\circ$ and \etacut, 
and $60^\circ<-\Delta\phi<120^\circ$ and \etacut. ``TransMAX" and ``transMIN" each have an area in \etaphi\ space 
of $2\pi/3$.  The sum of  ``TransMAX" and ``transMIN" is the total ``transverse" region with area $4\pi/3$.
}
\label{snow_fig11}
\end{figure}

\begin{figure}[htbp]
\includegraphics[scale=0.6]{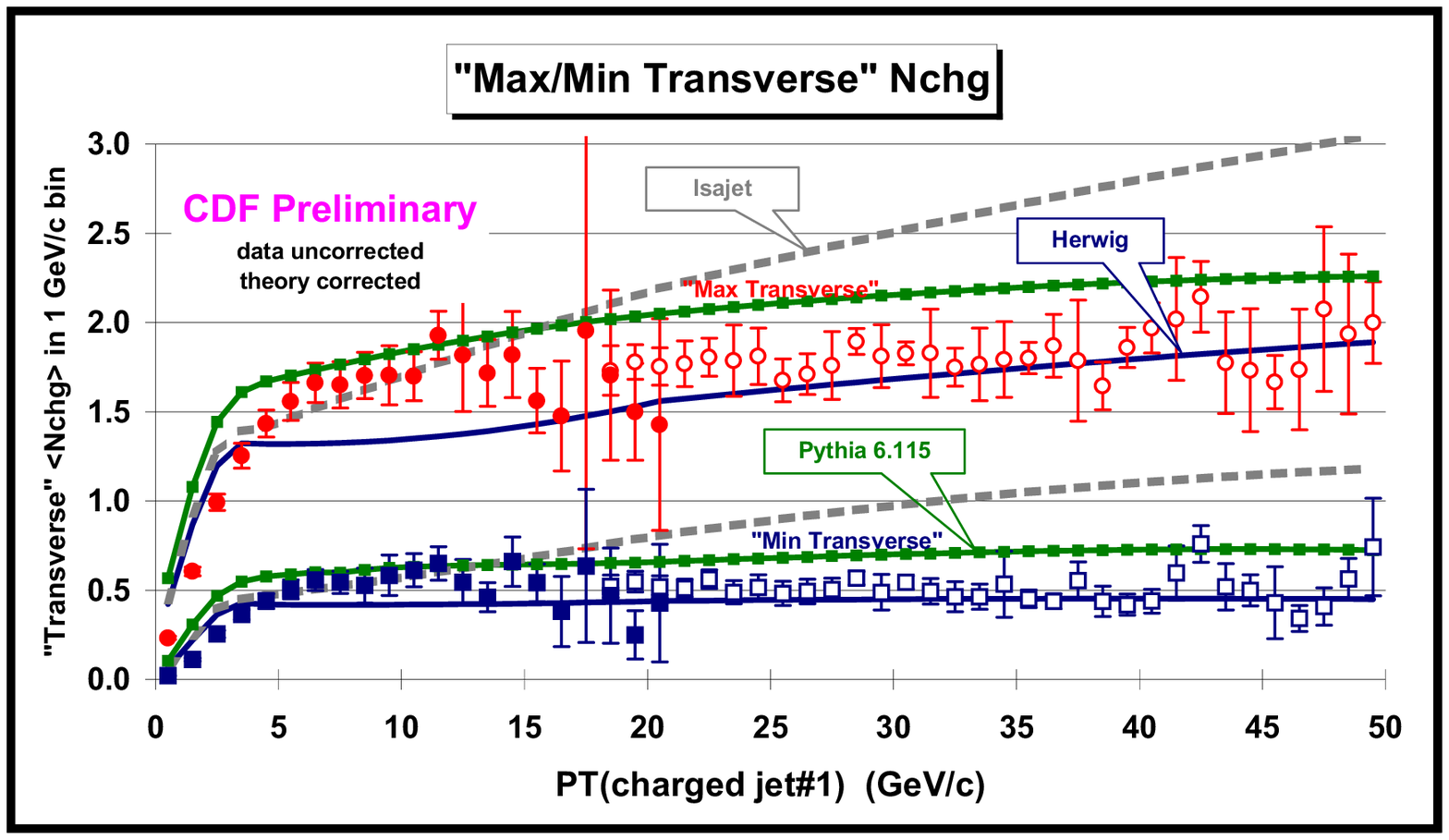}
\caption{Data on the average number of ``transMAX" and ``transMIN" charged particles (\ptcut, \etacut) as a function 
of the transverse momentum of the leading charged jet compared with the QCD Monte-Carlo predictions of 
HERWIG 5.9, ISAJET 7.32, and PYTHIA 6.115 with their default parameters and with \hardcut. The solid (open) points are 
the \MB\ (JET20) data. The theory curves are corrected for the track finding efficiency and have an error 
({\it statistical plus systematic}) of around $5\%$.
}
\label{snow_fig12}
\end{figure}

\begin{figure}[htbp]
\includegraphics[scale=0.6]{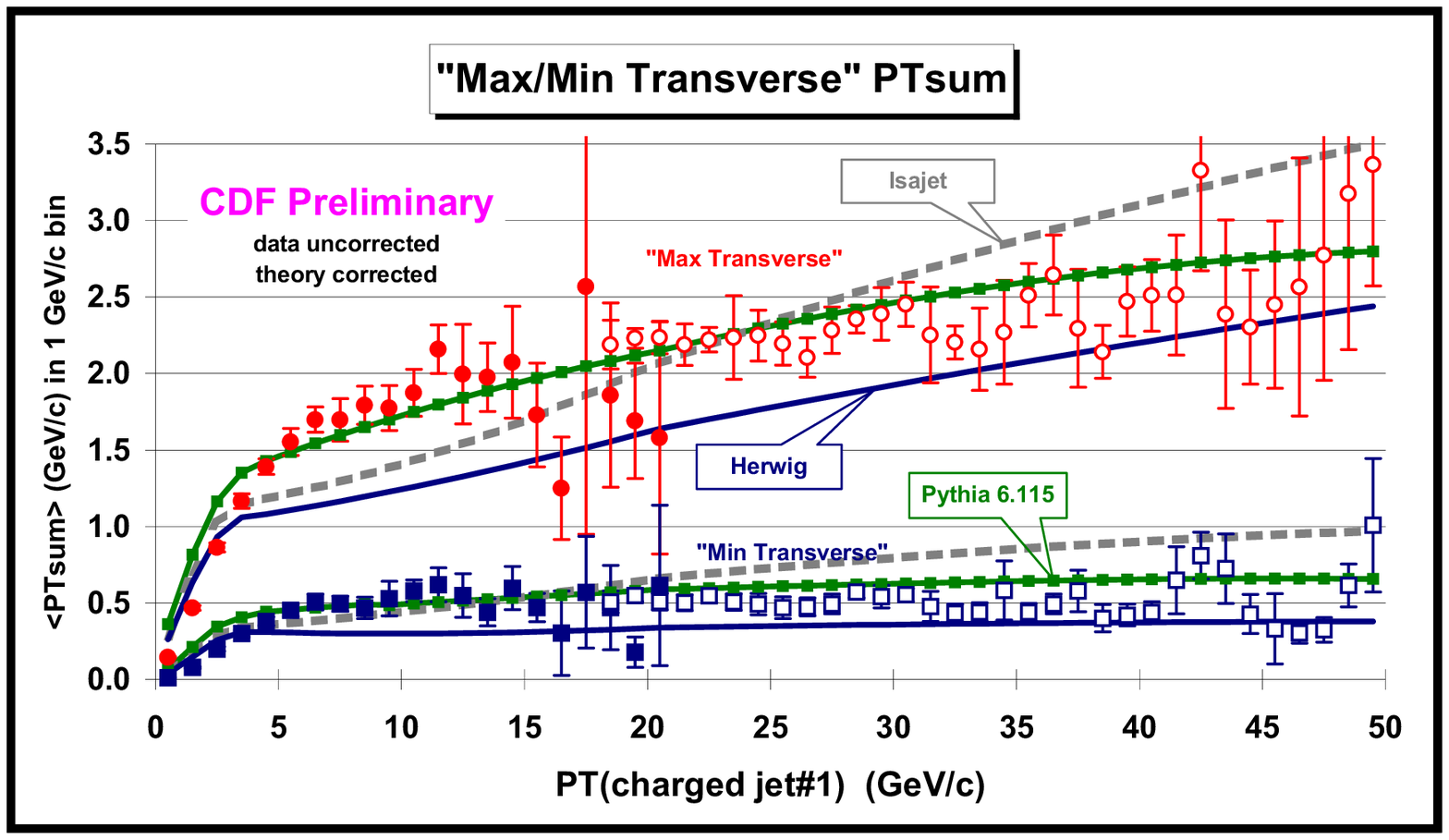}
\caption{Data on the average {\it scalar} \pt\ sum of the ``transMAX" and ``transMIN" charged 
particles (\ptcut, \etacut) as a function of the transverse momentum of the leading charged jet compared with the 
QCD Monte-Carlo predictions HERWIG 5.9, ISAJET 7.32, and PYTHIA 6.115 with their default parameters and with 
\hardcut. The solid (open) points are the \MB\ (JET20) data. The theory curves are corrected for the track 
finding efficiency and have an error ({\it statistical plus systematic}) of around $5\%$.
}
\label{snow_fig13}
\end{figure}

We expect the ``transverse" region to be composed predominately of particles that arise from the break-up of the 
beam and target and from initial and final-state radiation. This is clearly the case for the QCD Monte-Carlo models 
as can be seen in FIGS.~\ref{snow_fig7}-\ref{snow_fig9}, where the predictions for the ``transverse" region are divided into two categories: charged 
particles that arise from the break-up of the beam and target ({\it beam-beam remnants}), and charged particles that 
result 
from the outgoing jets plus initial and final-state radiation ({\it hard scattering component}). For PYTHIA the 
``beam-beam remnant" contribution includes contributions from multiple parton scattering.  It is interesting to see 
that in the QCD Monte-Carlo models it is the \BBR\ that are producing the approximately constant ``plateau".  
The contributions from initial-state and final-state radiation increase as \ptchj\ 
increases.  In fact, for ISAJET it is the sharp rise in the initial-state radiation component that is causing the 
disagreement with the data for \ptchj $>20\gevc$.  The hard scattering component of HERWIG and PYTHIA 
does not rise nearly as fast as the hard scattering component of ISAJET.  

There are two reasons why the hard scattering component of ISAJET is different from HERWIG and PYTHIA.  The 
first is due to different fragmentation schemes.  ISAJET uses independent fragmentation, which produces too many 
soft hadrons when partons begin to overlap.  The second difference arises from the way the QCD Monte-Carlo 
produce ``parton showers".  ISAJET uses a leading-log picture in which the partons within the shower are ordered 
according to their invariant mass.  Kinematics requires that the invariant mass of daughter partons be less than the 
invariant mass of the parent.  HERWIG and PYTHIA modify the leading-log picture to include ``color coherence 
effects" which leads to ``angle ordering" within the parton shower.  Angle ordering produces less high \pt\ radiation 
within a parton shower which is what is seen in FIG.~\ref{snow_fig10}.

Of course, the origin of an outgoing particle (``beam-beam remnant" or ``hard-scattering") is not an experimental 
observable.  Experimentally one cannot say where a given particle comes from.  However, we do know the origins of 
particles generated by the QCD Monte-Carlo models and FIGS.~\ref{snow_fig7}-\ref{snow_fig9} show the 
composition of the ``transverse" region as predicted by ISAJET, HERWIG, and PYTHIA. 

\begin{figure}[htbp]
\includegraphics[scale=0.6]{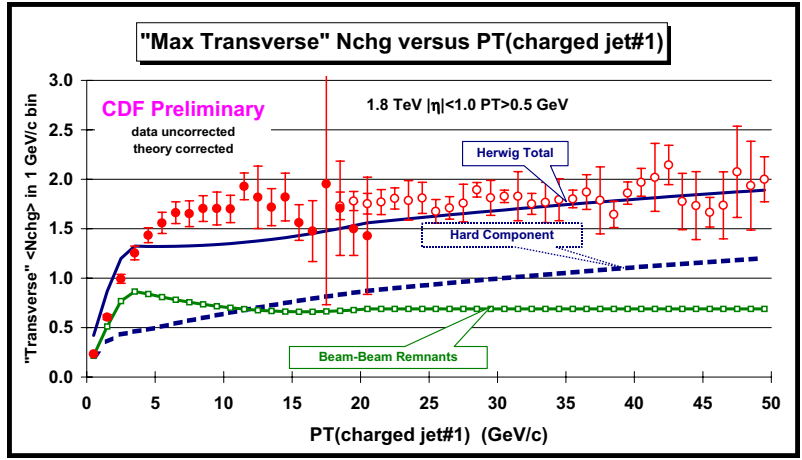}
\caption{Data on the average number of ``transMAX" charged particles (\ptcut, \etacut) as a function of the 
transverse momentum of the leading charged jet compared with the QCD Monte-Carlo predictions of 
HERWIG 5.9 (default parameters and \hardcut). The predictions of HERWIG are divided into two categories: 
charged particles that arise from the break-up of the beam and target ({\it beam-beam remnants}), and charged particles 
that result from the outgoing jets plus initial and final-state radiation ({\it hard scattering component}).  The theory 
curves are corrected for the track finding efficiency and have an error ({\it statistical plus systematic}) of 
around $5\%$.
}
\label{snow_fig14}
\end{figure}

\begin{figure}[htbp]
\includegraphics[scale=0.6]{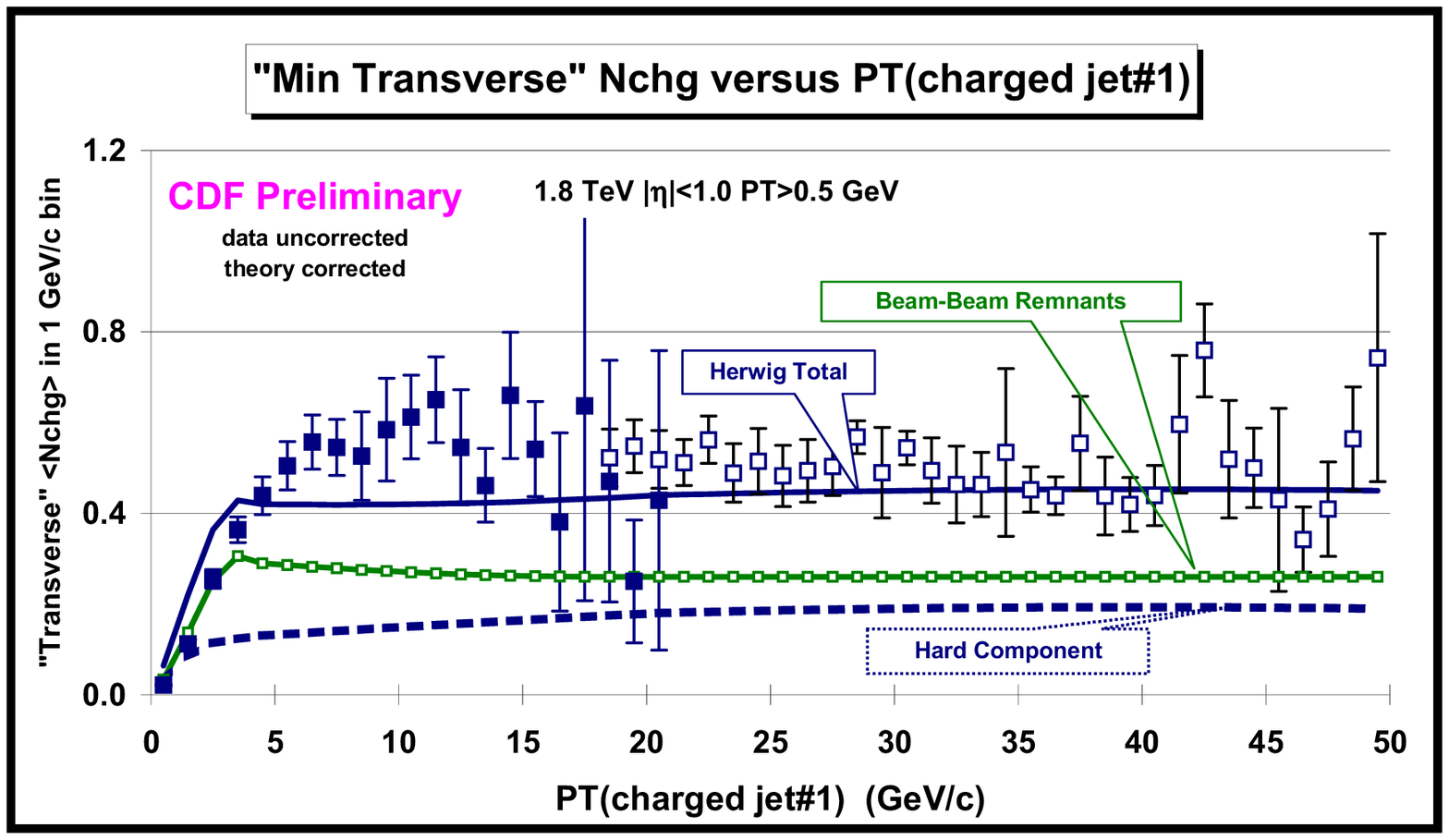}
\caption{Data on the average number of ``transMIN" charged particles (\ptcut, \etacut) as a function of the 
transverse momentum of the leading charged jet compared with the QCD Monte-Carlo predictions of 
HERWIG 5.9 (default parameters and \hardcut). The predictions of HERWIG are divided into two categories: 
charged particles that arise from the break-up of the beam and target ({\it beam-beam remnants}), and charged particles 
that result from the outgoing jets plus initial and final-state radiation ({\it hard scattering component}).  The theory 
curves are corrected for the track finding efficiency and have an error ({\it statistical plus systematic}) of 
around $5\%$.
}
\label{snow_fig15}
\end{figure}

\begin{figure}[htbp]
\includegraphics[scale=0.6]{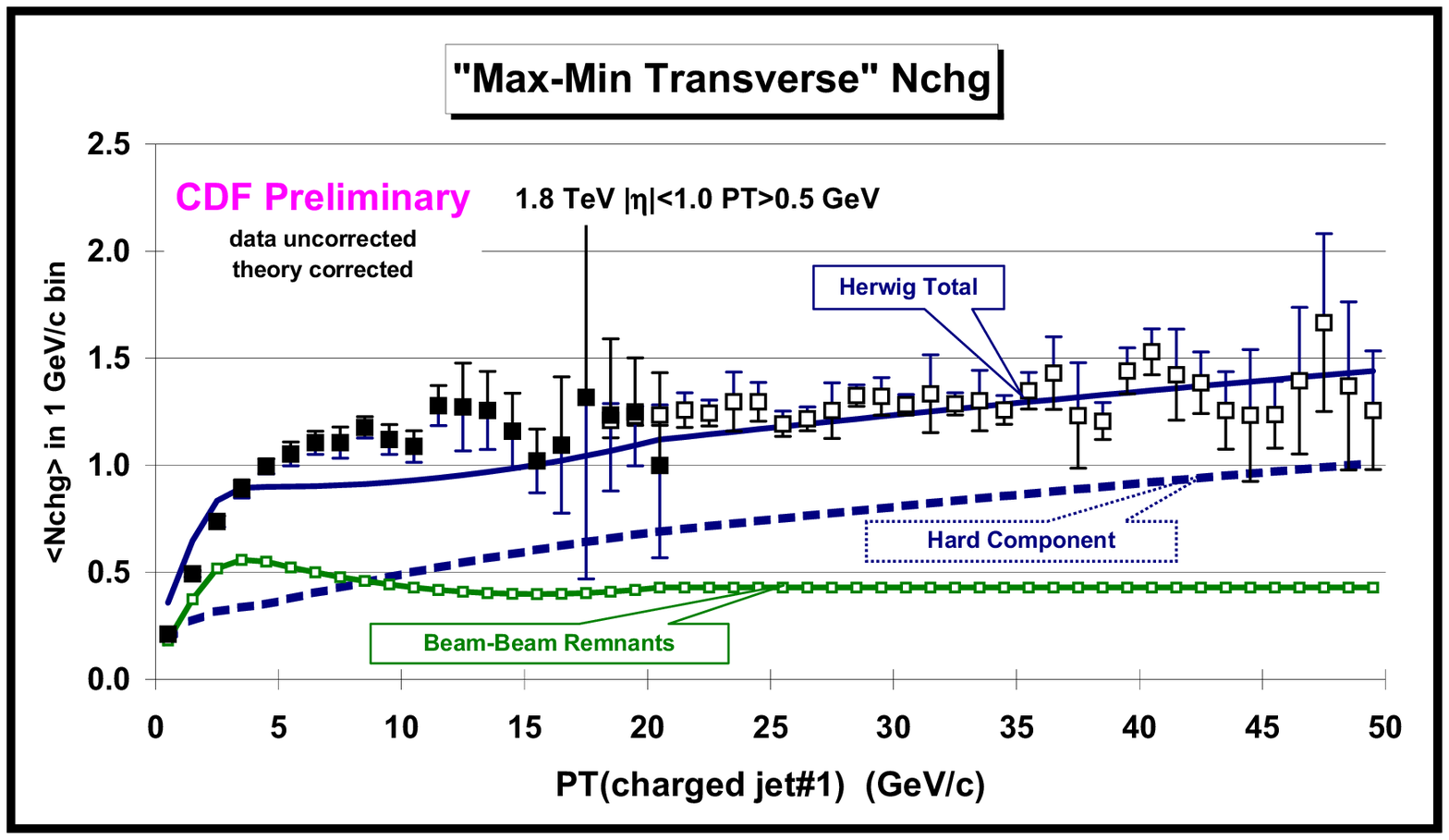}
\caption{Data on the average difference,``transMAX" minus ``transMIN", for the number of charged 
particles (\ptcut, \etacut) as a function of the 
transverse momentum of the leading charged jet compared with the QCD Monte-Carlo predictions of 
HERWIG 5.9 (default parameters and \hardcut). The predictions of HERWIG are divided into two categories: 
charged particles that arise from the break-up of the beam and target ({\it beam-beam remnants}), and charged particles 
that result from the outgoing jets plus initial and final-state radiation ({\it hard scattering component}).  The theory 
curves are corrected for the track finding efficiency and have an error ({\it statistical plus systematic}) of 
around $5\%$.
}
\label{snow_fig16}
\end{figure}

\section{Maximum and Minimum ``Transverse" Regions}

We now break up the ``transverse" region into two pieces.  As illustrated in FIG.~\ref{snow_fig11}, on an event 
by event basis, we 
define ``transMAX" (``transMIN") to be the maximum (minimum) of the two ``transverse" pieces, 
$60^\circ<\Delta\phi<120^\circ$, \etacut, and $60^\circ<-\Delta\phi<120^\circ$, \etacut.  Each has an area 
in \etaphi\ space of $2\pi/3$ and what we previously referred 
to as the ``transverse" region is the sum of  ``transMAX" and ``transMIN".  One expects that ``transMAX"  will pick 
up more of the initial and final state radiation while ``transMIN" should be more sensitive to the ``beam-beam 
remnant" component of the \UE.  Furthermore, one expects that the ``beam-beam remnant" component 
will nearly cancel in the difference, ``transMAX" minus ``transMIN".  If this is true then the difference, ``transMAX" 
minus ``transMIN", would be more sensitive to the ``hard scattering" component (\ie initial and final-state radiation).  
I believe that this idea was first proposed by Bryan Webber and then implemented in a paper by Jon Pumplin \cite{pumplin}.

FIG.~\ref{snow_fig12} and FIG.~\ref{snow_fig13} show the data on the \aveN\ and \avePT, respectively, for 
the``transMAX" and ``transMIN" region as a function of the \ptchj\ compared with QCD Monte-Carlo predictions of 
HERWIG, ISAJET, and PYTHIA with their default parameters and \hardcut.   FIG.~\ref{snow_fig14}, FIG.~\ref{snow_fig15}, 
and FIG.~\ref{snow_fig16} show the data 
on \aveN\  for ``transMAX",  ``transMIN", and the difference ``transMAX" minus ``transMIN", respectively, 
compared with QCD Monte-Carlo predictions of HERWIG.   The predictions of HERWIG are divided into two 
categories: charged particles that arise from the break-up of the beam and target ({\it beam-beam remnants}), 
and charged 
particles that result from the outgoing jets plus initial and final-state radiation ({\it hard scattering component}).  
It is 
clear from these plots that in the QCD Monte-Carlo models the ``transMAX" is more sensitive to the ``hard scattering 
component" of the \UE\ while ``transMIN" is more sensitive to the \BBR, especially 
at large values of \ptchj.   For example, for HERWIG at \ptchj $=40\gevc$ the hard scattering 
component makes up $62\%$ of the ``transMAX" \aveN\ with $38\%$ coming from the \BBR.  On the 
other hand, the hard scattering component makes up only $42\%$ of the ``transMIN" \aveN\ with $58\%$ coming from 
the \BBR\ at \ptchj $=40\gevc$.  Taking difference between ``tansMAX" and ``transMIN" does 
not completely remove the ``beam-beam remnant" component, but reduces it to only about $32\%$ at \ptchj $=40\gevc$.

\begin{figure}[htbp]
\includegraphics[scale=0.4]{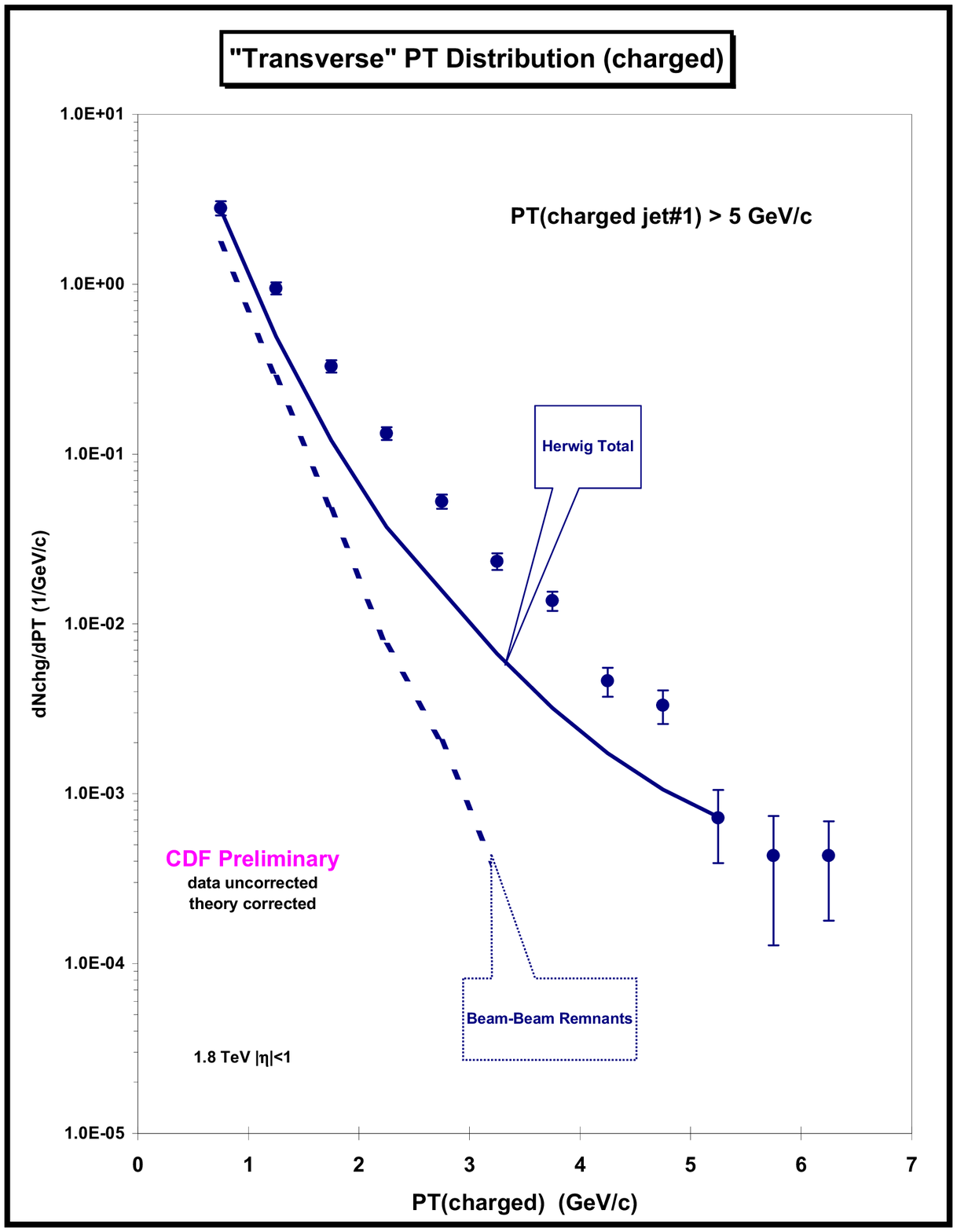}
\caption{Data on the transverse momentum distribution of charged particles (\ptcut, \etacut) in the ``transverse" region 
for \ptchj $> 5\gevc$ compared with the QCD Monte-Carlo model predictions of HERWIG 5.9 
(default parameters and \hardcut).  The theory curves are corrected for the track finding efficiency and have an 
error ({\it statistical plus systematic}) of around $5\%$. The solid curve is the total (``hard scattering" plus 
\BBR) and the dashed curve shows the contribution arising from the break-up of the beam 
particles (\BBR).
}
\label{snow_fig18}
\end{figure}

\begin{figure}[htbp]
\includegraphics[scale=0.4]{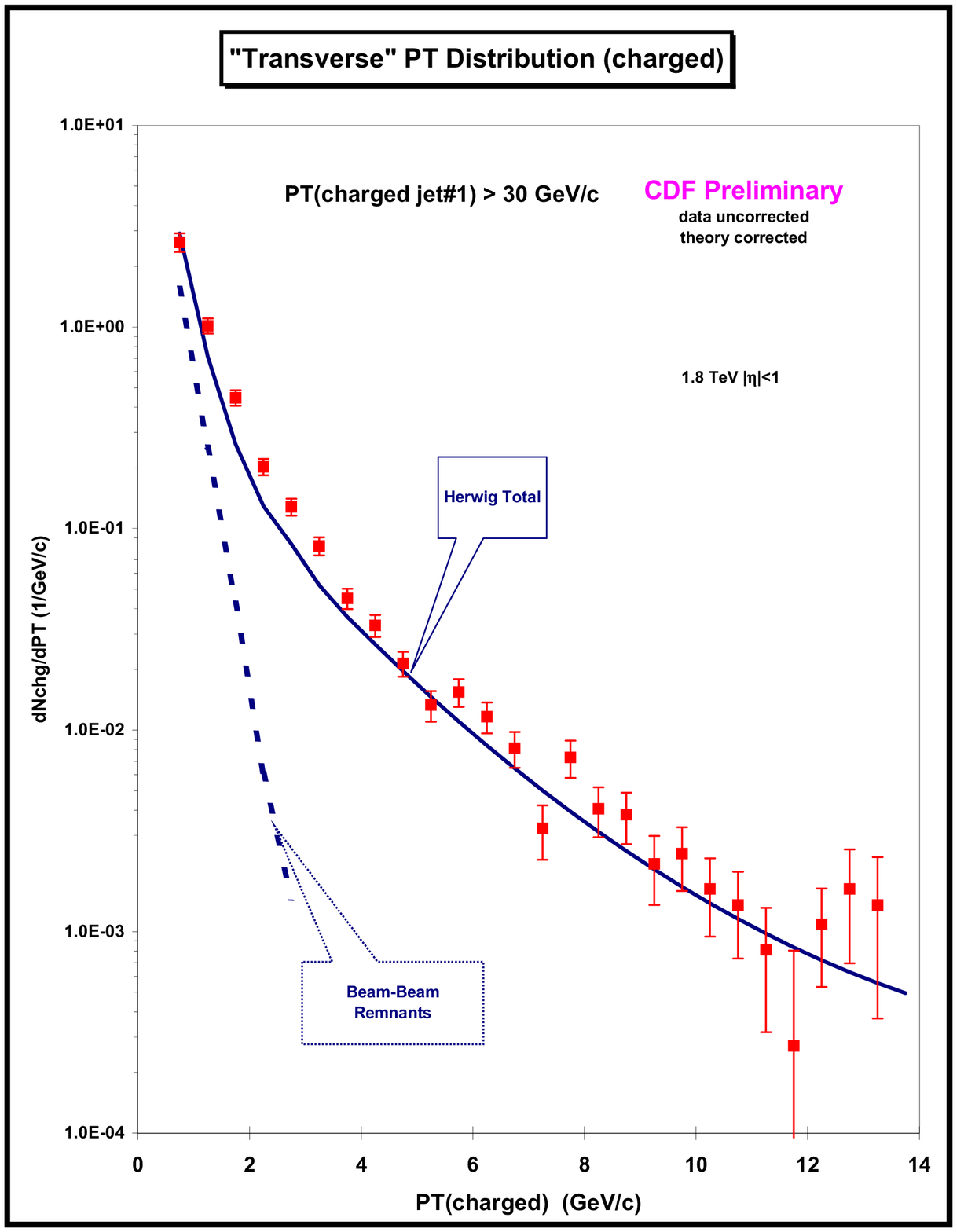}
\caption{Data on the transverse momentum distribution of charged particles (\ptcut, \etacut) in the ``transverse" region 
for \ptchj $> 30\gevc$ compared with the QCD Monte-Carlo model predictions of HERWIG 5.9 
(default parameters and \hardcut).  The theory curves are corrected for the track finding efficiency and have an 
error ({\it statistical plus systematic}) of around $5\%$. The solid curve is the total (``hard scattering" plus 
\BBR) and the dashed curve shows the contribution arising from the break-up of the beam 
particles (\BBR).
}
\label{snow_fig19}
\end{figure}

\section{The Transverse Momentum Distribution in the ``Transverse" Region}

FIG.~\ref{snow_fig17} shows the data on the transverse momentum distribution of charged particles in the ``transverse" 
region for \ptchj $> 2\gevc$, $5\gevc$, and $30\gevc$. Each point corresponds to $dN_{\rm chg}/dp_T$ and the integral 
of the 
distribution gives the average number of charged particles in the ``transverse" region, \aveN, shown in 
FIG.~\ref{snow_fig5}.  FIG.~\ref{snow_fig5} shows only mean values, while FIG.~\ref{snow_fig17} shows the 
distribution from which the mean is computed.  The 
data are compared with the QCD hard scattering Monte-Carlo models predictions HERWIG.  Since these 
distributions fall off sharply as \pt\ increases, it is essentially only the first few points at low \pt\ that determine the 
mean.  The approximately constant plateau seen in FIG.~\ref{snow_fig5} is a result of the low \pt\ points in 
FIG.~\ref{snow_fig17} not changing 
much as \ptchj\ changes.  However, the high \pt\ points do increase considerably as \ptchj\ increases. This 
effect cannot be seen by simply examining the average number of  ``transverse" particles.  FIG.~\ref{snow_fig17} 
shows the growth 
of the ``hard scattering component" at large \pt\ in the "transverse region" (\ie three or more hard scattering jets).
For the QCD Monte-Carlo models the \pt\ distribution in the ``transverse" region, at low values of \ptchj, is 
dominated by the ``beam-beam remnant" contribution with very little ``hard scattering" component.  This can be seen 
in FIG.~\ref{snow_fig18} which shows both the ``beam-beam remnant" component together with the total overall 
prediction of 
HERWIG for \ptchj $>5\gevc$.  For the QCD Monte-Carlo models the \pt\ distribution in the ``transverse" 
region, at low values of \ptchj, measures directly the \pt\ distribution of the \BBR.  Both 
ISAJET and HERWIG have the wrong \pt\ dependence in the ``transverse" region due to a ``beam-beam remnant" 
component that falls off too rapidly as \pt\ increases.  It is, of course, understandable that the Monte-Carlo models 
might be slightly off on the parameterization of the \BBR.  This component cannot be calculated 
from perturbation theory and must be determined from data.

\begin{figure}[htbp]
\includegraphics[scale=0.6]{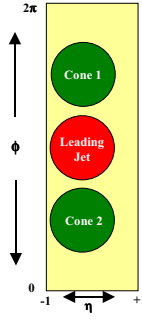}
\caption{Illustration of ``transverse cones" with radius in \etaphi\ space of $R = 0.7$ which are located at the 
same pseudorapicity as the leading jet but with azimuthal angle $\Delta\phi=+90^\circ$ and  $\Delta\phi=-90^\circ$ 
relative to the leading jet.  Each``transverse cone" has an area in \etaphi\ space of $\pi R^2 = 0.49\pi$.
}
\label{snow_fig20}
\end{figure}

\begin{figure}[htbp]
\includegraphics[scale=0.8]{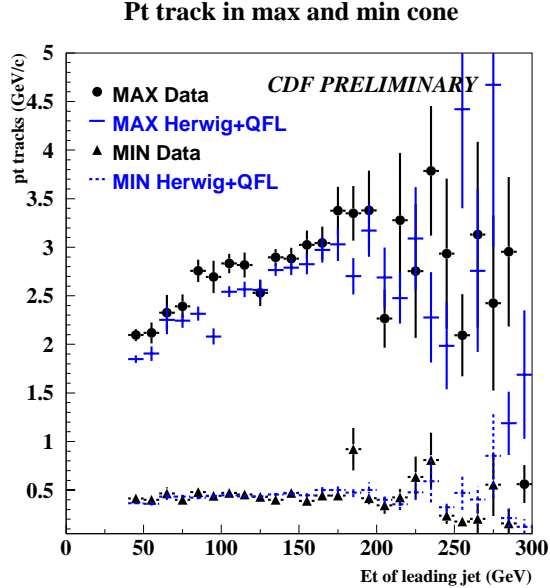}
\caption{Data on the average {\it scalar} \pt\ sum of charged particles ($p_T > 0.4\gevc$, \etacut) within the 
maximum (MAX) and minimum (MIN) ``transverse cones" defined in FIG.~\ref{snow_fig20} versus the transverse energy of 
the leading ({\it highest $E_T$}) ``calorimeter jet" compared with the QCD Monte-Carlo model predictions 
HERWIG 5.9 ({\it default parameters}).  
}
\label{snow_fig21}
\end{figure}

FIG.~\ref{snow_fig19} shows both the ``beam-beam remnant" component together with the overall prediction of HERWIG for 
\ptchj $>30\gevc$.  Here the QCD Monte-Carlo models predict a large ``hard scattering" component 
corresponding to the production of more than two large \pt\ jets. HERWIG, ISAJET, and PYTHIA all do well at 
describing the high \pt\ tail of this distribution.  

\section{``Transverse Regions" Versus ``Transverse Cones"}

In a complementary CDF analysis Valeria Tano \cite{tano} has studied the \UE\ in hard scattering processes 
by defining ``transverse cones" instead of ``transverse regions".  As illustrated in FIG.~\ref{snow_fig20}, 
the ``transverse cones" (with 
radius in \etaphi\ space of $R = 0.7$) are located at the same pseudo-rapidity as the leading jet but 
with azimuthal angle $\Delta\phi=+90^\circ$ and $\Delta\phi=-90^\circ$ relative to the leading ``jet".    
In the cone analysis the ``jet" is a ``calorimeter jet" ({\it charged plus neutrals}) defined using 
the standard CDF cluster algorithm.  Maximum (MAX) and minimum (MIN) 
``transverse" cones are determined, on an event-by-event basis, similar to the ``transMAX" and ``transMIN" regions 
described in Section III.  Each ``transverse cone" has an area in \etaphi\ space of $\pi R^2 = 0.49\pi$ 
(compared with $0.67\pi$).
FIG.~\ref{snow_fig21} shows data at $1.8\tev$ on the average {\it scalar} \pt\ sum of charged particles 
($p_T > 0.4\gevc$, \etacut) within the 
MAX and MIN ``transverse cones" versus the transverse energy of the leading ({\it highest $E_T$}) ``calorimeter jet" 
compared with the QCD hard scattering Monte-Carlo models predictions HERWIG.  The ``transverse cone" analysis 
covers the range $50 < E_T({\rm calorimeter\ jet}\#1) < 300\gev$, while the ``transverse region" analysis examines only 
charged particles and covers the range $0 <$\ptchj $<50\gevc$.  One cannot directly compare the two 
analysis, but if one scales the low $E_T({\rm jet}\#1)$ points in FIG.~\ref{snow_fig21} by the ratio of 
areas $0.67\pi/0.49\pi=1.36$, one gets 
approximate agreement with the high \ptchj\ points in FIG.~\ref{snow_fig13}.  Both FIG.~\ref{snow_fig13} 
and FIG.~\ref{snow_fig21} indicate that the 
\avePT\ of charged particles generated by HERWIG is slightly too small.  Together the two CDF analyses give us a 
good handle on the \UE\ in hard scattering processes.

\section{Tuning PYTHIA to Fit the ``Underlying Event"}

Now that we have constructed collider observables that are sensitive to the \UE\ we would like to tune 
the multiple parton interaction parameters of PYTHIA to fit the data.  There are many tunable parameters.  Here we 
consider only the parameters given in Table~\ref{snow_table1}.  The default values of the parameters are given in 
Table~\ref{snow_table2}.  Note that 
the PYTHIA default values sometimes change as the version changes \cite{new_pythia}.

\begin{table}[htbp]
\caption{PYTHIA multiple parton scattering parameters.}
\label{snow_table1}
\begin{tabular}{||c|c|l||}
\hline\hline
{\bf Parameter} & {\bf Value} & {\bf Description} \\
\hline\hline
MSTP(81) &	$0$  &  Multiple-Parton Scattering off \\
\hline
	   &  $1$  &  Multiple-Parton Scattering on  \\
\hline
MSTP(82) &	$1$  &  Multiple interactions assuming the same probability, \\
& & with an abrupt cut-off $P_T{\rm min}$=PARP(81) \\
\hline
	   &  $3$  &  Multiple interactions assuming a varying impact parameter \\
& & and a hadronic matter overlap consistent with a \\
& & single Gaussian matter distribution, with a smooth turn-off $P_{T0}$=PARP(82) \\
\hline
	   &  $4$  &  Multiple interactions assuming a varying impact parameter \\
& & and a hadronic matter overlap consistent with a \\
& & double Gaussian matter distribution (governed by PARP(83) and PARP(84)) \\
& & with a smooth turn-off $P_{T0}$=PARP(82) \\
\hline\hline
\end{tabular}
\end{table}

\begin{table}[htbp]
\caption{Default values for some of the multiple parton scattering parameters of PYTHIA.}
\label{snow_table2}
\begin{tabular}{||c|c|c||}
\hline\hline
{\bf Parameter} & {\bf PYTHIA 6.115} & {\bf PYTHIA 6.125} \\
\hline\hline
MSTP(81) &	$1$  &  $1$ \\
\hline
MSTP(82) &  $1$  &  $1$  \\
\hline
PARP(81) &	$1.4\gevc$  & $1.9\gevc$ \\
\hline
PARP(82) &	$1.55\gevc$  & $2.1\gevc$ \\
\hline
PARP(83) &	$0.5$  & $0.5$ \\
\hline
PARP(84) &	$0.2$  & $0.2$ \\
\hline\hline
\end{tabular}
\end{table}

\begin{figure}[htbp]
\includegraphics[scale=0.6]{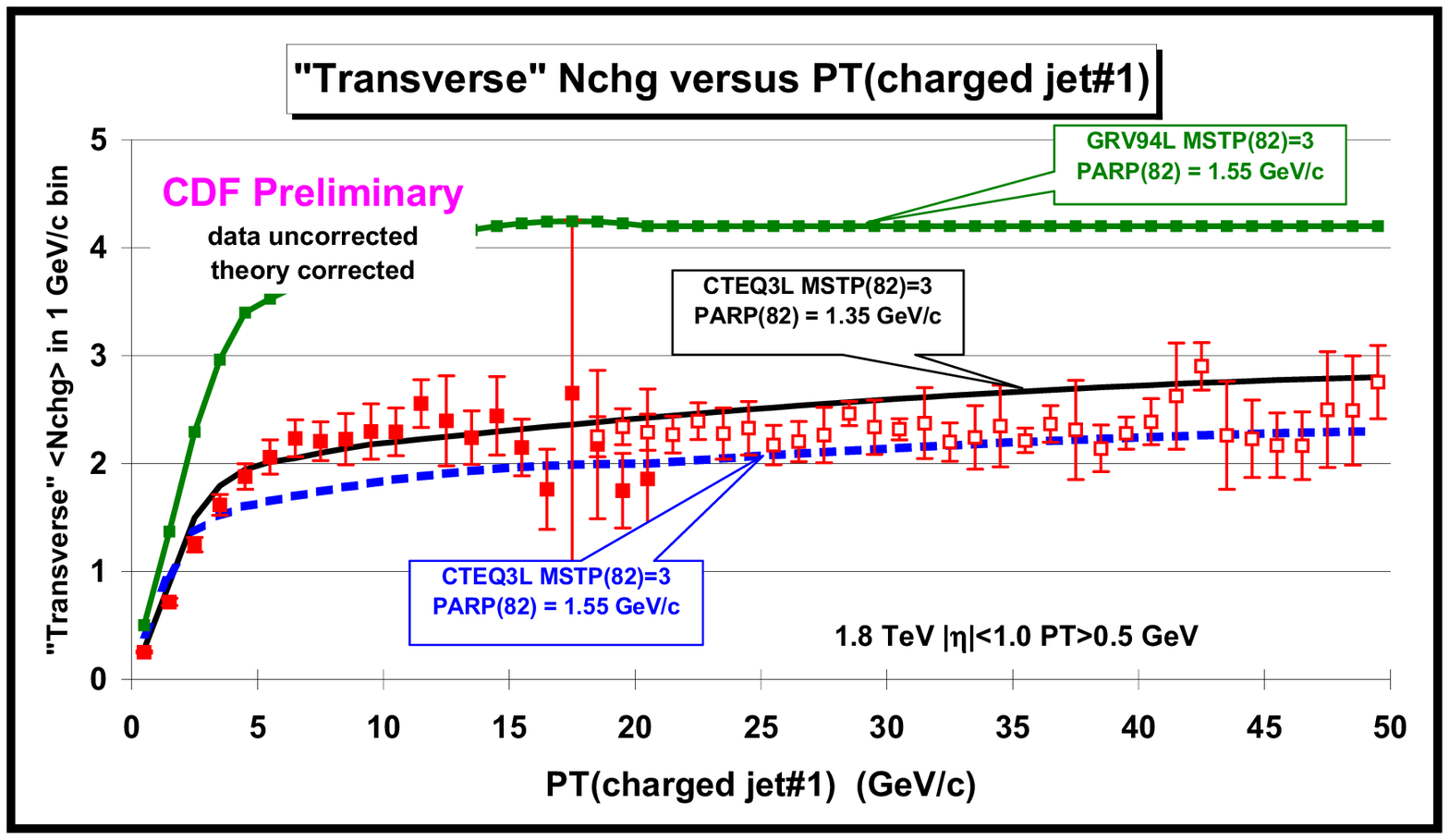}
\caption{Data on the average number of charged particles (\ptcut, \etacut) in the ``transverse" region 
as a function of the transverse momentum of the leading charged jet compared with the QCD Monte-Carlo predictions of 
PYTHIA 6.115 with different structure functions and different multiple parton interaction parameters and 
with \hardzero. The theory curves are corrected for the track finding efficiency and have an 
error ({\it statistical plus systematic}) of around $5\%$.
}
\label{snow_fig22}
\end{figure}

\begin{figure}[htbp]
\includegraphics[scale=0.6]{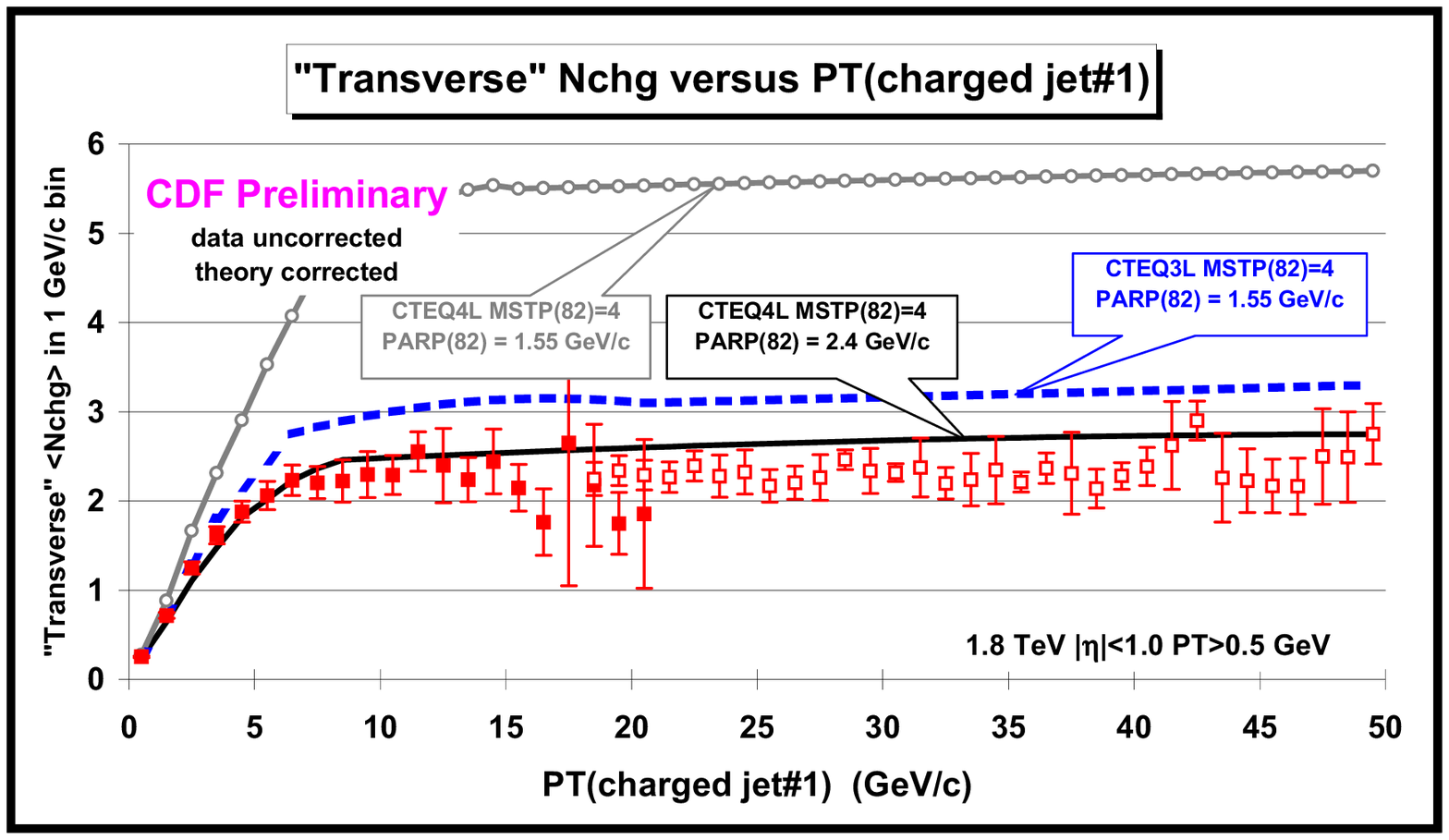}
\caption{Data on the average number of charged particles (\ptcut, \etacut) in the ``transverse" region 
as a function of the transverse momentum of the leading charged jet compared with the QCD Monte-Carlo predictions of 
PYTHIA 6.115 with different structure functions and different multiple parton interaction parameters and 
with \hardzero. The theory curves are corrected for the track finding efficiency and have an 
error ({\it statistical plus systematic}) of around $5\%$.
}
\label{snow_fig23}
\end{figure}

FIG.~\ref{snow_fig22} and FIG.~\ref{snow_fig23} show data on the average number of charged 
particles in the ``transverse" region compared with 
the QCD Monte-Carlo predictions of PYTHIA 6.115 with different structure functions and different multiple parton 
interaction parameters and with \hardzero.  For PYTHIA the amount of multiple parton scattering depends 
on the parton distribution functions (\ie the structure functions) and hence the number of particles produced in the 
``transverse" region (\ie the \UE) changes if one changes the structure functions.  HERWIG and 
ISAJET do not include multiple parton scattering and for them the number of particles in the ``transverse" is 
essentially independent of the choice of structure functions.

\begin{figure}[htbp]
\includegraphics[scale=0.6]{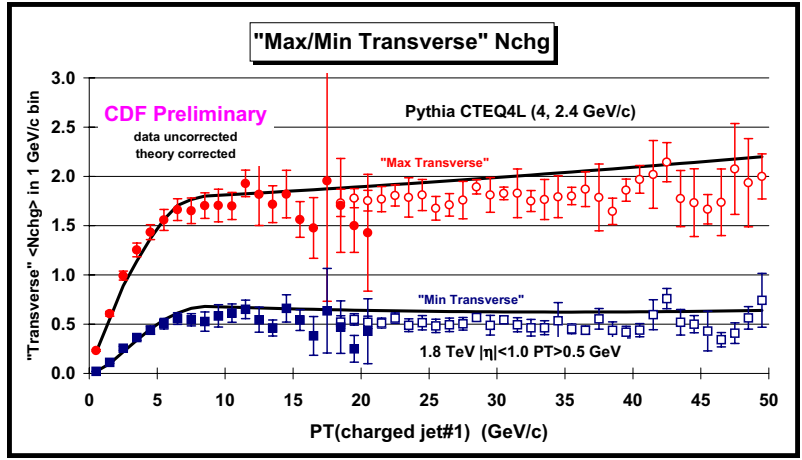}
\caption{Data on the average number of ``transMAX" and ``transMIN" charged particles (\ptcut, \etacut) as a function 
of the transverse momentum of the leading charged jet defined compared with the QCD Monte-Carlo predictions of 
PYTHIA 6.115 ({\it tuned version}, CTEQ4L, MSTP(82) = 4, PARP(82) = $1.4\gevc$, \hardzero). The theory curves are 
corrected for the track finding efficiency and have an error ({\it statistical plus systematic}) of around $5\%$. 
}
\label{snow_fig24}
\end{figure}

\begin{figure}[htbp]
\includegraphics[scale=0.6]{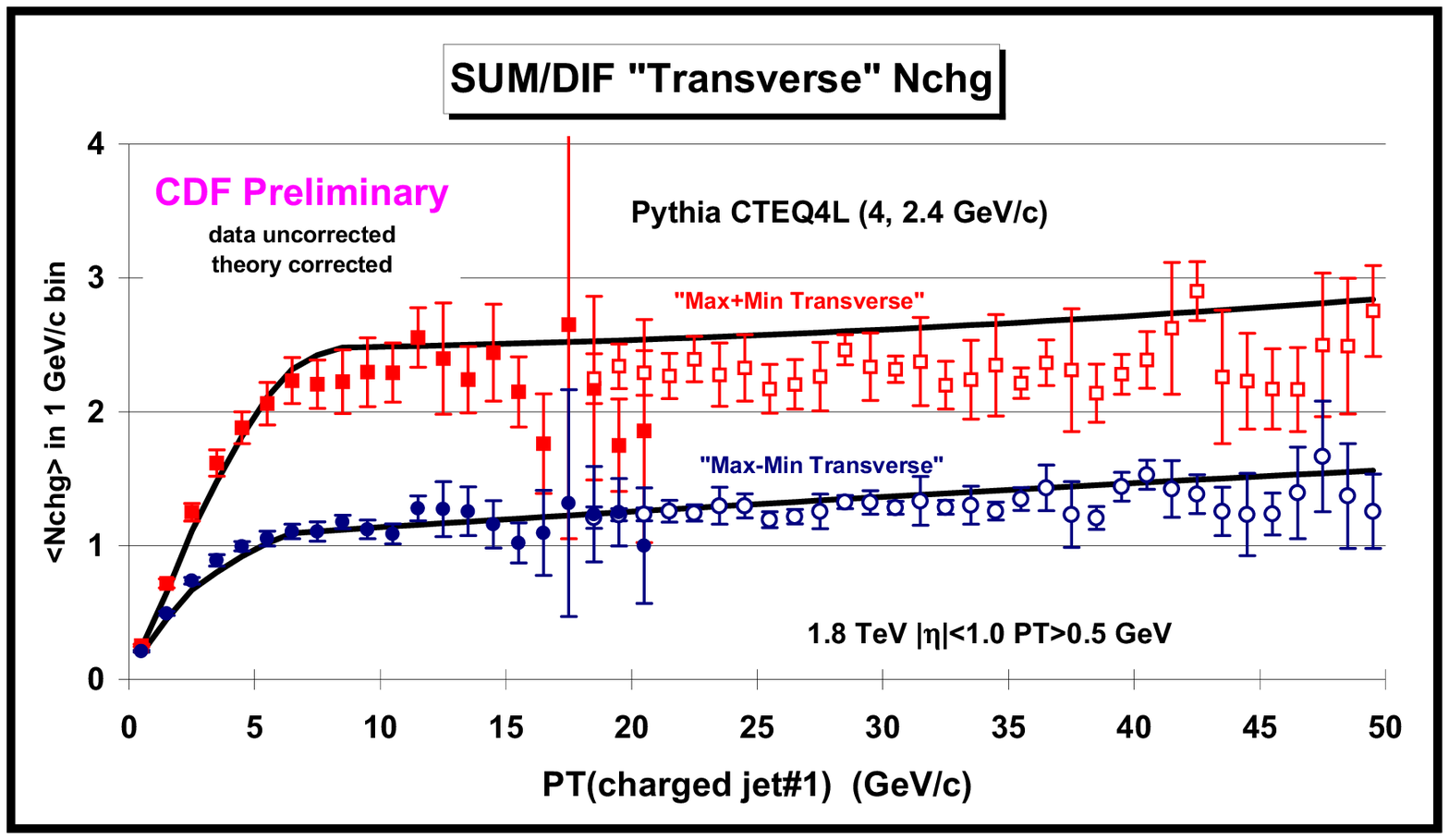}
\caption{Data on the average sum,``transMAX" plus ``transMIN", and difference, ``transMAX" minus ``transMIN" 
for the number of charged particles (\ptcut, \etacut) as a function 
of the transverse momentum of the leading charged jet defined compared with the QCD Monte-Carlo predictions of 
PYTHIA 6.115 ({\it tuned version}, CTEQ4L, MSTP(82) = 4, PARP(82) = $1.4\gevc$, \hardzero). The theory curves are 
corrected for the track finding efficiency and have an error ({\it statistical plus systematic}) of around $5\%$. 
}
\label{snow_fig25}
\end{figure}

\begin{figure}[htbp]
\includegraphics[scale=0.6]{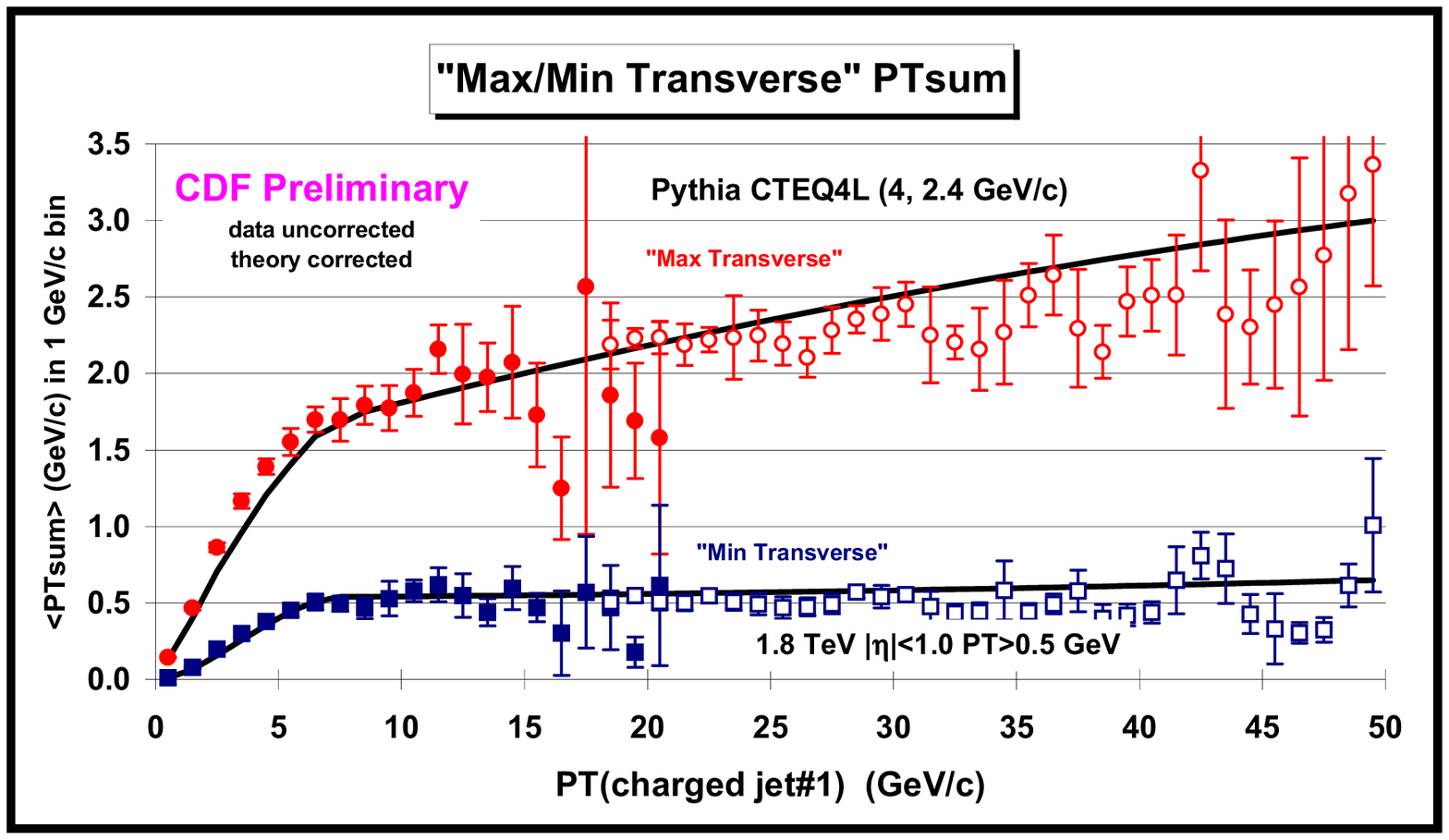}
\caption{Data on the average {\it scalar} \pt\ sum of ``transMAX" and ``transMIN" charged particles (\ptcut, \etacut) 
as a function of the transverse momentum of the leading charged jet defined compared with the QCD Monte-Carlo 
predictions of PYTHIA 6.115 ({\it tuned version}, CTEQ4L, MSTP(82) = 4, PARP(82) = $1.4\gevc$, \hardzero). The theory 
curves are corrected for the track finding efficiency and have an error ({\it statistical plus systematic}) of 
around $5\%$. 
}
\label{snow_fig26}
\end{figure}

\begin{figure}[htbp]
\includegraphics[scale=0.6]{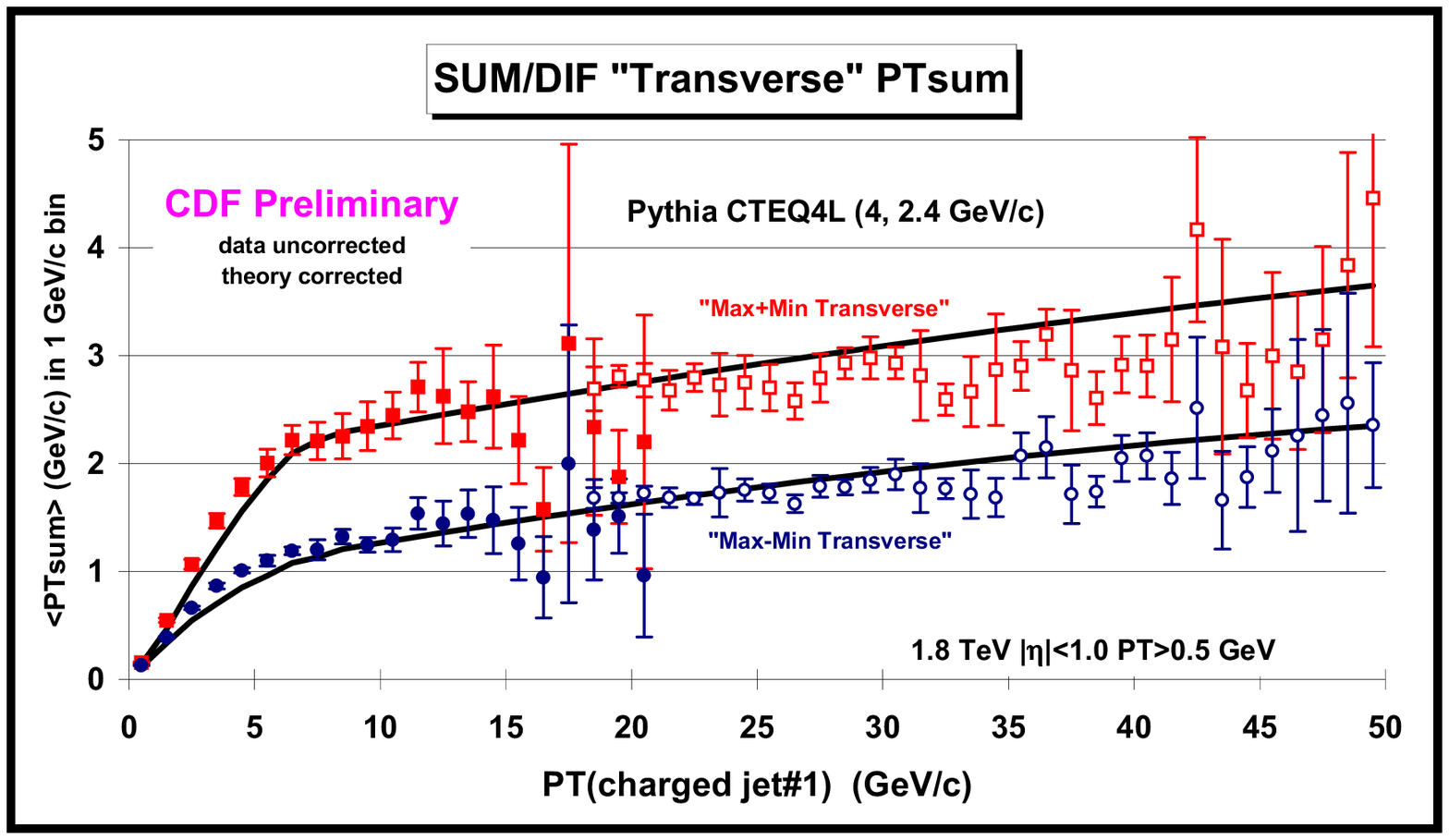}
\caption{Data on the average sum, ``transMAX" plus ``transMIN", and difference, ``transMAX" minus ``transMIN" 
for the {\it scalar} \pt\ sum of charged particles (\ptcut, \etacut) as a function 
of the transverse momentum of the leading charged jet defined compared with the QCD Monte-Carlo predictions of 
PYTHIA 6.115 ({\it tuned version}, CTEQ4L, MSTP(82) = 4, PARP(82) = $1.4\gevc$, \hardzero). The theory curves are 
corrected for the track finding efficiency and have an error ({\it statistical plus systematic}) of around $5\%$. 
}
\label{snow_fig27}
\end{figure}

\begin{figure}[htbp]
\includegraphics[scale=0.4]{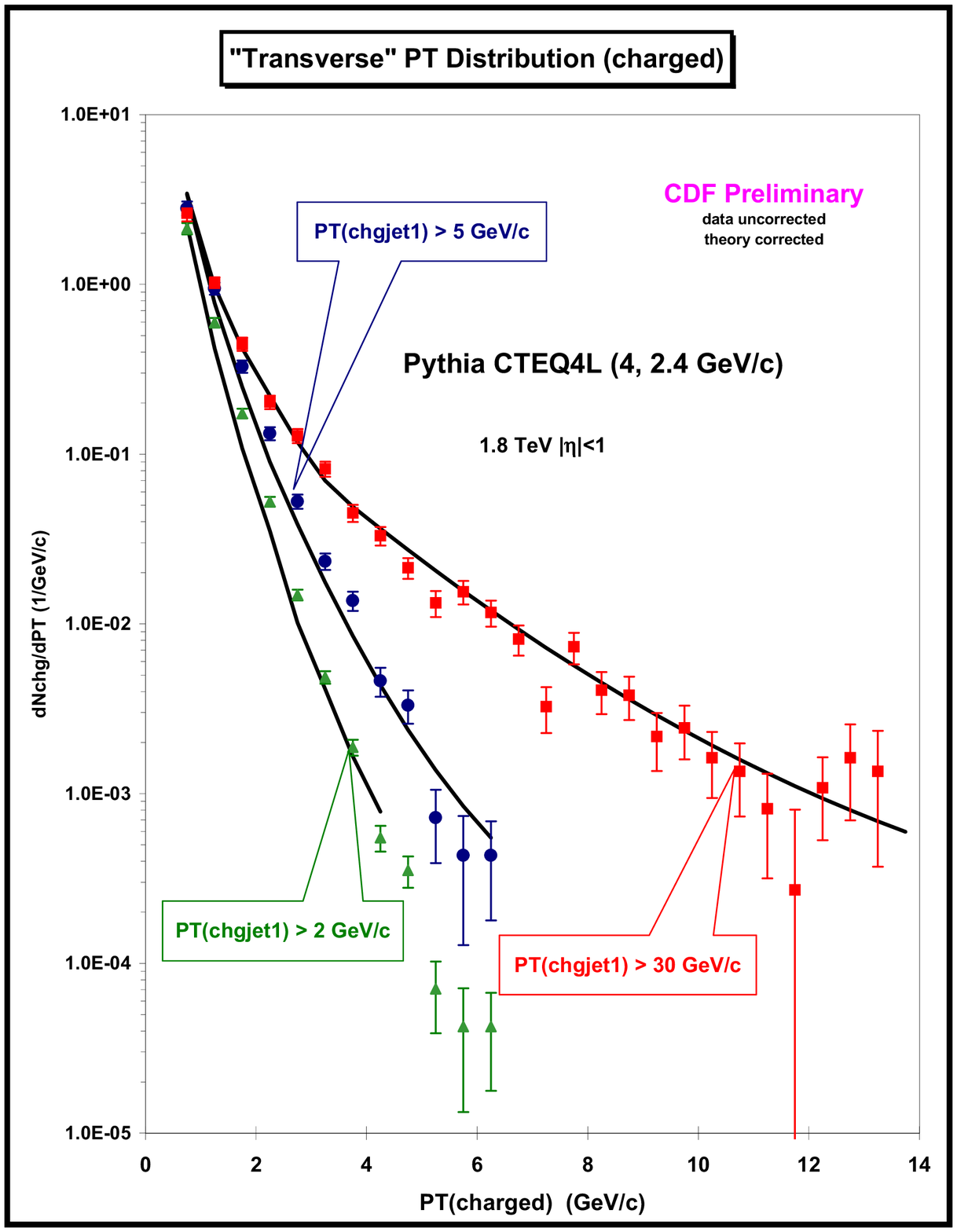}
\caption{Data on the transverse momentum distribution of charged particles (\ptcut, \etacut) in the ``transverse" region 
for \ptchj $>2\gevc$, $5\gevc$, and $30\gevc$, where chgjet\#1 is the leading charged particle jet. Each point 
corresponds to $dN_{chg}/dp_T$ and the integral of the distribution gives the average number of charged particles 
in the transverse region, $\langle N_{chg}({\rm transverse})\rangle$. The data are compared with the QCD 
Monte-Carlo model predictions of PYTHIA 6.115 ({\it tuned version}, CTEQ4L, MSTP(82) = 4, PARP(82) = $1.4\gevc$, \hardzero).  
The theory curves are corrected for the track finding efficiency and have an error ({\it statistical plus systematic}) 
of around $5\%$.
}
\label{snow_fig28}
\end{figure}

\begin{figure}[htbp]
\includegraphics[scale=0.4]{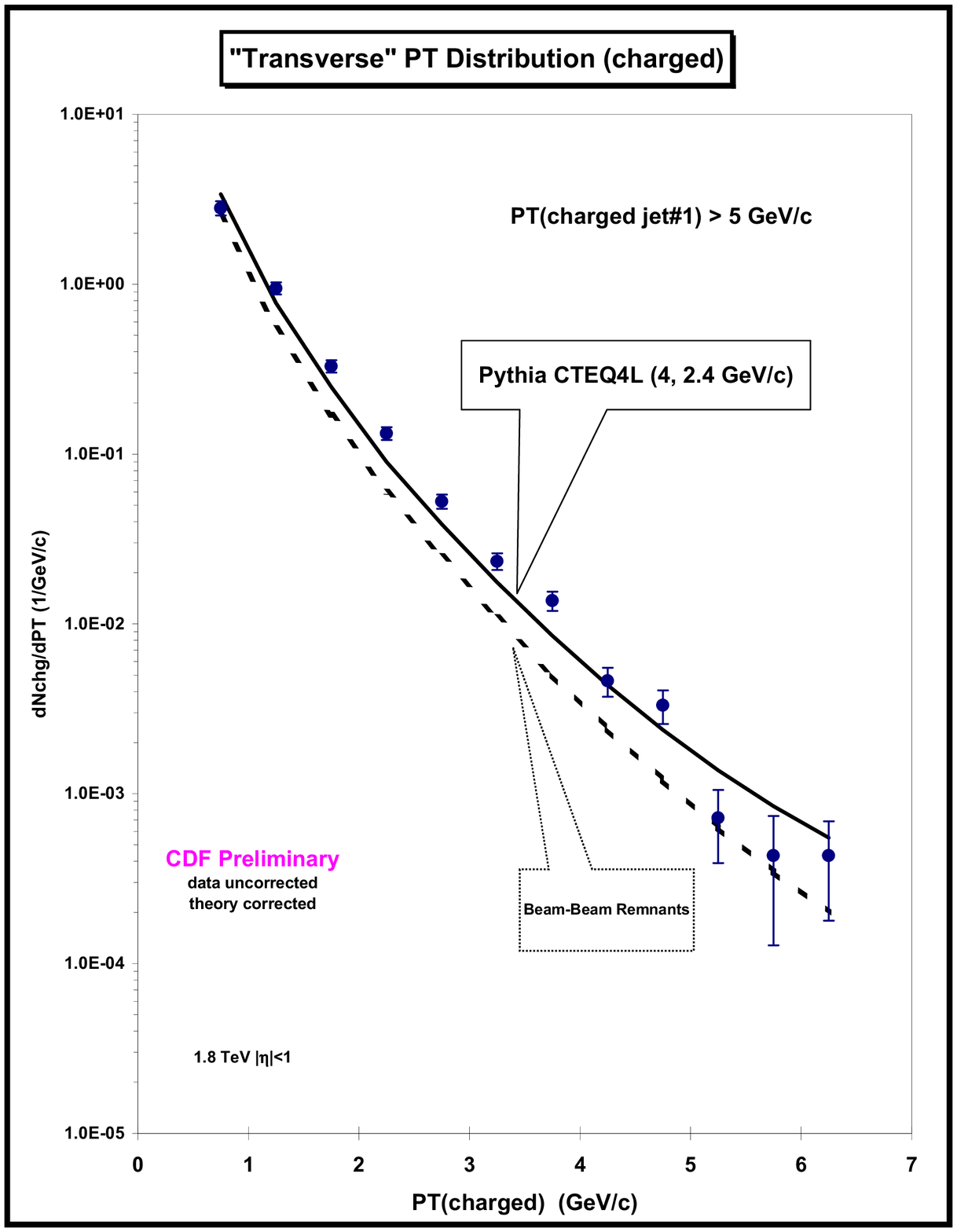}
\caption{Data on the transverse momentum distribution of charged particles (\ptcut, \etacut) in the ``transverse" region 
for \ptchj $>5\gevc$, compared with the QCD Monte-Carlo model predictions of PYTHIA 6.115 
({\it tuned version}, CTEQ4L, MSTP(82) = 4, PARP(82) = $1.4\gevc$, \hardzero).  The theory curves are corrected for 
the track finding efficiency and have an error ({\it statistical plus systematic}) of around $5\%$. The solid curve 
is the total (``hard scattering" plus \BBR) and the dashed curve shows the contribution arising from the 
break-up of the beam particles (\BBR). For PYTHIA the \BBR\ include contributions from multiple parton 
scattering. 
}
\label{snow_fig29}
\end{figure}

FIGS.~\ref{snow_fig24}-~\ref{snow_fig29} show the results of a ``tuned" version of 
PYTHIA 6.115 with MSTP(82) = $4$ and PARP(82) = $2.4\gevc$ 
using the CTEQ4L structure functions.  One must first choose a structure function and then tune the multiple parton 
scattering parameters for that structure function.  In generating the PYTHIA curves in 
FIGS.~\ref{snow_fig24}-~\ref{snow_fig29}  we have taken 
\hardzero.  In general the perturbative $2$-to-$2$ parton scattering subprocesses diverge as \pthard\ goes to 
zero.  PYTHIA regulates these divergences using the same cut-off parameters that are used to regulate the multiple 
parton scattering cross section (see Table~\ref{snow_table1}).  This allows for the possibility of using PYTHIA to simultaneously 
describe both ``soft" and ``hard" collisions.  Most of the CDF \MB\ events are ``soft", with less than $3\%$ of the 
events having \ptchj $>5\gevc$.  There is no clear separation between ``soft" and ``hard" collisions, but 
roughly speaking \ptchj $<2\gevc$ corresponds to ``soft" \MB\ collisions and demanding \ptchj $>5\gevc$ 
assures a ``hard" collision.  FIGS.~\ref{snow_fig24}-~\ref{snow_fig27} show that the ``tuned" version of 
PYTHIA with \hardzero\ describes fairly well the transition between ``soft" and ``hard" collisions.  
The QCD Monte-Carlo models with \hardcut\ cannot describe the data for \ptchj $<3\gevc$ 
(see FIG.~\ref{snow_fig5} and FIG.~\ref{snow_fig6}), whereas PYTHIA 
with \hardzero\ seems to do a good job on the ``transverse" observables as  \ptchj\ goes to zero.

FIG.~\ref{snow_fig28} shows the data on the transverse momentum distribution of charged particles in the ``transverse" 
region compared with the ``tuned" version of PYTHIA 6.115 (CTEQ4L, MSTP(82) = $4$, PARP(82) = $2.4\gevc$).  The fit is 
not perfect, but it is much better than the HERWIG prediction shown in FIG.~\ref{snow_fig17}.  Multiple parton 
scattering produces 
more large \pt\ particles in the ``transverse" region, which is what is needed to fit the data.  
As seen in FIG.~\ref{snow_fig29}, the \pt\ distribution in the ``transverse" region, at low values of \ptchj, 
for the ``tuned" version of PYTHIA is also 
dominated by the ``beam-beam remnant" contribution as is the case for HERWIG (see FIG.~\ref{snow_fig18}).  However, for 
PYTHIA the ``beam-beam remnant" component includes contributions from multiple parton scattering, which results 
in a less steep \pt\ distribution. 

\section{Summary and Conclusions}

The \UE\ in a hard scattering process is a complicated and interesting object which involves aspects of 
both non-perturbative and perturbative QCD.  Studying the ``transMAX" and ``transMIN" pieces of the ``transverse" 
region provides additional information not contained in the sum.  In the QCD Monte-Carlo models the various 
components that make up the \UE\ are weighted differently in ``transMAX" and ``transMIN" terms.  
The ``transMAX" term preferentially selects the ``hard component" of the \UE\ ({\it outgoing jets plus 
initial and final-state radiation}) while the ``transMIN" term preferentially selects the ``beam-beam remnant" 
component.  Unfortunately one cannot cleanly isolate a single component of the \UE\ since all 
components contribute to both  ``transMAX", ``transMIN", and to the difference.  However, requiring the 
Monte-Carlo models to fit both ``transMAX" and ``transMIN" (or the sum and difference) puts additional constraints 
on the way the generators model the \UE.  

ISAJET ({\it with independent fragmentation}) produces too many ({\it soft}) particles in the \UE\ with the 
wrong dependence on \ptchj. HERWIG and PYTHIA modify the leading-log picture to include ``color coherence 
effects" which leads to ``angle ordering" within the parton shower and they do a better job describing the ``underlying 
event".  Both ISAJET and HERWIG have the too steep of a \pt\ dependence of the ``beam-beam remnant" component 
of the \UE\ and hence do not have enough \BBR\ with \ptcut.  PYTHIA 
with multiple parton scattering does the best job at fitting the data.

The increased activity in the \UE\ of a hard scattering over that observed in  ``soft" collisions cannot be 
explained solely by initial-state radiation. Multiple parton interactions provide a natural way of explaining the 
increased activity in the \UE\ in a hard scattering.  A hard scattering is more likely to occur when 
the ``hard cores" of the beam hadrons overlap and this is also when the probability of a multiple parton interaction is 
greatest.  For a soft grazing collision the probability of a multiple parton interaction is small. However, multiple 
parton interactions are very sensitive to the parton structure functions (PDF).  You must first decide on a particular 
PDF and then tune the multiple parton interactions to fit the data.  

One should not take the ``tuned" version of PYTHIA 6.115 (CTEQ4L, MSTP(82) = $4$, PARP(82) = $2.4\gevc$) 
presented here too seriously.  It is encouraging that it describes fairly well the "transverse" region over the 
range $0<$\ptchj$<50\gevc$ including the transition from ``soft" to ``hard" collisions.  However, it is still not 
quite right.  For example, it does not reproduce very well the multiplicity distribution of ``soft" collisions.  
More work needs to be done in tuning the Monte-Carlo models \cite{butterworth}.  In addition, more work needs to be 
done before one 
can say that the multiple parton interaction approach is correct.   HERWIG without multiple parton scattering is 
not that far off the data.  Maybe we simply need to change and improve the way the Monte-Carlo models handle the 
``beam-beam remnant" component.

%\begin{acknowledgments}
%\end{acknowledgments}

% \bibliography{your bib file}

\begin{thebibliography}{99}
\bibitem{new_pythia} The latest versions of PYTHIA (6.120 and higher) include additional parameters that allow one to adjust the 
energy dependence of multiple parton interactions.
\end{thebibliography}

\end{document}